%% file: arxiv_quantumsurvey.tex
\pdfoutput=1

\documentclass[10pt]{article}

\usepackage{authblk}                
\usepackage[square,numbers]{natbib} 

\usepackage{amsmath,amssymb}                

\input{common_settings.tex}

\input{header.tex}

\input{generated_files/collection_statistics.tex}
\input{generated_files/RQ_results.tex}

\usepackage{geometry}
\title{
\input{title.tex}
}

\author{Man Zhang}
\author{Yuechen Li}
\author{Tao Yue\thanks{Corresponding author}}
\author{Kai-Yuan Cai}

\affil{Beihang University}

\date{}

\begin{document}

\maketitle

\begin{abstract}
\input{abstract}
\end{abstract}

{\bf Keywords}: \input{keywords.tex}

\input{content.tex}


\bibliographystyle{ACM-Reference-Format} 

%

\bibliography{qo_papers}

\newpage
\input{appendix.tex}

\end{document}

%% file: common_settings.tex
\usepackage{listings}    
\usepackage{graphicx}    
\usepackage{subcaption}  
\usepackage{algorithm}   
\usepackage{algorithmic} 
\usepackage{booktabs}    
\usepackage{hyperref}    
\usepackage{xurl}        
\usepackage[htt]{hyphenat} 
\usepackage{microtype} 
\DisableLigatures{}    

\usepackage{caption}
\usepackage{setspace}
\usepackage{multirow}
\usepackage{enumerate}
\usepackage{pdflscape}
\usepackage{pgfplots} 

\usepackage{xcolor}
\definecolor{codegreen}{rgb}{0.25,0.5,0.35}
\definecolor{codegray}{rgb}{0.5,0.5,0.5}
\definecolor{codepurple}{rgb}{0.6,0,0}
\definecolor{backcolour}{rgb}{0.95,0.95,0.92}
\definecolor{colorstring}{rgb}{0.5,0,0.35}
\definecolor{rltred}{rgb}{0.5,0,0}
\definecolor{rltgreen}{rgb}{0,0.5,0}
\definecolor{rltblue}{rgb}{0,0,0.5}
\definecolor{DarkGreen}{rgb}{0.00,0.60,0.00}
\definecolor{ScarletRed}{rgb}{0.80,0.00,0.00}
\definecolor{blizzardblue}{rgb}{0.67, 0.9, 0.93}
\definecolor{green-yellow}{rgb}{0.68, 1.0, 0.18}
\definecolor{dkgreen}{rgb}{0,0.6,0}
\definecolor{gray}{rgb}{0.5,0.5,0.5}
\definecolor{mauve}{rgb}{0.58,0,0.82}
\definecolor{lightgrey}{rgb}{0.90,0.90,0.90}
\definecolor{grey}{gray}{0.75}
\definecolor{light-gray}{gray}{0.80}

\lstdefinestyle{mystyle}{
    escapechar=©, 
    language=Python,
	backgroundcolor=\color{backcolour},
    basicstyle=\footnotesize\ttfamily,
   	identifierstyle=\footnotesize\ttfamily,
	commentstyle=\color{codegreen},
	keywordstyle=\color{colorstring}\bfseries,
	morekeywords={OR, AND},
	numberstyle=\ttfamily\color{codegray},
	stringstyle=\ttfamily\color{DarkGreen},
	breakatwhitespace=false,
	breaklines=true,
	captionpos=b,
	keepspaces=true,
	numbers=left, 
	numbersep=2pt,
	showspaces=false,
	showstringspaces=false,
	showtabs=false,
	tabsize=2
}
\usepackage{braket}		
\usepackage{mathtools}	

\usepackage{pifont} 
\newcommand{\markyes}{\ding{51}}  
\newcommand{\markno}{\ding{55}}   

\usepackage{xspace}

\usepackage{tcolorbox}
\newtcolorbox{resultsbox}[1][]
{
	colframe=gray!100, 
	colback=white!100, 
	coltitle=white,
	title=#1 
}

\newenvironment{results}[1][]{
	\begin{resultsbox}[#1]
	}{
	\end{resultsbox}
}

\usepackage{ifthen}
\newboolean{showcomments}
\setboolean{showcomments}{true} 

\ifthenelse{\boolean{showcomments}}{
	\newcommand{\nbc}[3]{
		{\colorbox{#3}{\bfseries\sffamily\scriptsize\textcolor{white}{#1}}}
		{\textcolor{#3}{\sf\small$\langle$\textit{#2}$\rangle$}}}
	
}{
	\newcommand{\nbc}[3]{}

}



%% file: header.tex
\newcommand{\rqAYear}{\textbf{RQ1:} How many primary studies have been published over the years?\xspace}
\newcommand{\rqAVenue}{\textbf{RQ2:} What are  the publication venues of all primary studies?\xspace}
\newcommand{\rqAPublicationType}{\textbf{RQ3:} What types of publications do the primary studies belong?\xspace}

\newcommand{\rqBSEActivity}{\textbf{RQ4:} What types of SE activities are targeted by quantum-related optimization solutions?\xspace}
\newcommand{\rqBSEProblem}{\textbf{RQ5:}  Which SE problems are solved by quantum-related optimization solutions?\xspace}
\newcommand{\rqBSEReformulation}{\textbf{RQ6:}  How are SE problems reformulated into optimization problems?\xspace 
}

\newcommand{\rqCQOType}{\textbf{RQ7:} What kinds of quantum optimization solutions are proposed for solving SE problems?\xspace} 
\newcommand{\rqCQOInspired}{\textbf{RQ8:} What inspires quantum-inspired optimization algorithms?\xspace} 
\newcommand{\rqCQOComputing}{\textbf{RQ9:} How are quantum optimization solutions applied to solve SE problems?\xspace}

\newcommand{\rqEClaimed}{\textbf{RQ10:} What are research challenges claimed in the primary studies?\xspace} 
\newcommand{\rqEOpen}{\textbf{RQ11:} What are open research challenges in applying quantum-based solutions to solve SE problems?\xspace}

\newcommand{\totalNumRQs}{11\xspace}

\newcommand{\totalNumInitalSearch}{2083\xspace}

\newcommand{\initialSearchDate}{January-13-2025\xspace}
\newcommand{\snowballingSearchDate}{April-25-2025\xspace}

\newcommand{\swebok}{SWEBOK\xspace}

%% file: generated_files/collection_statistics.tex
\newcommand{\dbIEEE}{IEEE\emph{Xplore}\xspace}
\newcommand{\dbACM}{ACM\xspace}
\newcommand{\dbScopus}{Scopus\xspace}
\newcommand{\dbWiley}{Wiley\xspace}
\newcommand{\dbWoS}{Web of Science\xspace}
\newcommand{\dbSpringer}{Springer Nature Link\xspace}

\newcommand{\initialtotal}{2083\xspace}

\newcommand{\finalselected}{77\xspace}
\newcommand{\finalsnowballing}{38\xspace}
\newcommand{\finalquery}{39\xspace}

\newcommand{\paperfirst}{jhaveri2023cloning}
\newcommand{\papersecond}{ammermann2024quantum}
\newcommand{\paperthird}{gan2024research}
\newcommand{\paperfourth}{ren2024dynamic}
\newcommand{\paperfifth}{schonberger2023digitalannealing}
\newcommand{\papersixth}{barletta2024quantum}
\newcommand{\paperseventh}{groppe2021optimizing}
\newcommand{\papertenth}{bajaj2022test}
\newcommand{\papereleventh}{mandal2024evaluating}
\newcommand{\papertwelfth}{deepalakshmi2022optimized}
\newcommand{\paperthirteenth}{wang2024quantum}
\newcommand{\paperfourteenth}{dornala2023quantum}
\newcommand{\papersixteenth}{zhang2017test}
\newcommand{\papertwentieth}{guo2023effective}
\newcommand{\papertwentysecond}{chen2020quantum}
\newcommand{\papertwentythird}{rani2023novel}
\newcommand{\papertwentyfourth}{kiruthiga2019enriched}
\newcommand{\papertwentyfifth}{serrano2024minimizing}
\newcommand{\papertwentysixth}{wu2020hybrid}
\newcommand{\papertwentyseventh}{bettonte2022quantum}
\newcommand{\papertwentyeighth}{kumari2016comparing}
\newcommand{\papertwentyninth}{liu2024quantum}
\newcommand{\paperthirtieth}{jin2021cross}
\newcommand{\paperthirtyfirst}{zhang2021new}
\newcommand{\paperthirtysecond}{jin2016parameter}
\newcommand{\paperthirtythird}{bajaj2022improved}
\newcommand{\paperthirtyfourth}{guo2022synergic}
\newcommand{\paperthirtyfifth}{tong2023can}
\newcommand{\paperthirtysixth}{xiong2016optimized}
\newcommand{\paperthirtyseventh}{hussein2021quantum}
\newcommand{\paperthirtyeighth}{bajaj2022Sensorstest}
\newcommand{\paperthirtyninth}{xiong2016virtual}
\newcommand{\paperfortieth}{fi13100260}
\newcommand{\paperfortyfirst}{jain2023quantum}
\newcommand{\paperfortysecond}{deng2022research}
\newcommand{\paperfortythird}{muniswamy2024joint}
\newcommand{\paperfortyfourth}{pathak2023opposition}
\newcommand{\paperfortyfifth}{bhatia2020quantum}
\newcommand{\paperfortyeighth}{ou2023quantum}
\newcommand{\paperfiftieth}{saxena2024constrained}
\newcommand{\paperfiftysecond}{uotila2025left}
\newcommand{\paperfiftythird}{trummer2015multiple}
\newcommand{\paperfiftyfourth}{nayak2023constructing}
\newcommand{\paperfiftyfifth}{barletta2022quantum}
\newcommand{\paperfiftysixth}{caivano2022quantum}
\newcommand{\paperfiftyeighth}{fankhauser2023multiple}
\newcommand{\papersixtieth}{trummer2024leveraging}
\newcommand{\papersixtyfirst}{schonberger2023quantum}
\newcommand{\papersixtythird}{trovato2025reformulating}
\newcommand{\papersixtyfourth}{wang2024test}
\newcommand{\papersixtyfifth}{miranskyy2022using}
\newcommand{\papersixtysixth}{wagner2019quantum}
\newcommand{\papersixtyseventh}{lou2018failure}
\newcommand{\papersixtyeighth}{blanco2024qiss}
\newcommand{\papersixtyninth}{barletta2023quantum}
\newcommand{\paperseventieth}{zhang2020intelligent}
\newcommand{\paperseventyfirst}{jin2015prediction}
\newcommand{\paperseventysecond}{balicki2021many}
\newcommand{\paperseventythird}{shen2021failure}
\newcommand{\paperseventyfourth}{kumar2025optimizing}
\newcommand{\paperseventyfifth}{guo2023tolerance}
\newcommand{\paperseventysixth}{guo2023memetic}
\newcommand{\paperseventyseventh}{hussein2020quantum}
\newcommand{\paperseventyninth}{roy2024quantum}
\newcommand{\papereightieth}{qiao2020novel}
\newcommand{\papereightysecond}{naik2025energy}
\newcommand{\papereightythird}{pathak2024cooperative}
\newcommand{\papereightyfourth}{urgelles2022multi}
\newcommand{\papereightyfifth}{liu2022hpcp}
\newcommand{\papereightysixth}{ahanger2022quantum}
\newcommand{\papereightyseventh}{emu2022resource}
\newcommand{\papereightyeighth}{emu2024warm}
\newcommand{\papereightyninth}{krishna2023optimal}
\newcommand{\paperninetieth}{niu2024performance}
\newcommand{\paperninetyfirst}{bhatia2019quantum}
\newcommand{\paperninetysecond}{li2019quantum}
\newcommand{\paperninetythird}{schonberger2023ready}


%% file: generated_files/RQ_results.tex
\newcommand{\RQfirstmaxNumYear}{22\xspace}
\newcommand{\RQfirstmaxYearfirst}{2023\xspace}
\newcommand{\RQfirstmaxYearsecond}{2024\xspace}
\newcommand{\RQsecondmaxNumConference}{5\xspace}
\newcommand{\RQsecondmaxConference}{\emph{IEEE International Conference on Quantum Computing and Engineering (QCE)}\xspace}

\newcommand{\RQsecondmaxJournalfirst}{\emph{Applied Soft Computing}\xspace}
\newcommand{\RQsecondmaxJournalsecond}{\emph{Cluster Computing}\xspace}

\newcommand{\RQthirdtotalNumConference}{21\xspace}
\newcommand{\RQthirdprecentConference}{27.27\%\xspace}
\newcommand{\RQthirdtotalNumJournal}{49\xspace}
\newcommand{\RQthirdprecentJournal}{63.64\%\xspace}
\newcommand{\RQthirdtotalNumOpenAccessArchive}{3\xspace}
\newcommand{\RQthirdprecentOpenAccessArchive}{3.90\%\xspace}
\newcommand{\RQthirdtotalNumWorkshop}{8\xspace}
\newcommand{\RQthirdprecentWorkshop}{10.39\%\xspace}
\newcommand{\RQthirdNumSEVenue}{15\xspace}
\newcommand{\RQthirdprecentSEVenue}{19.48\%\xspace}

\newcommand{\RQthirdprecentNonSEVenue}{80.52\%\xspace}
\newcommand{\RQfourthtopfirstSEActivityNum}{43\xspace}
\newcommand{\RQfourthtopfirstSEActivity}{\emph{software engineering operations}\xspace}
\newcommand{\RQfourthprecentTopfirstSEActivityNum}{55.84\%\xspace}

\newcommand{\RQfourthtopsecondSEActivity}{\emph{software testing}\xspace}
\newcommand{\RQfourthprecentTopsecondSEActivityNum}{20.78\%\xspace}

\newcommand{\RQfourthtopthirdSEActivity}{\emph{software quality}\xspace}
\newcommand{\RQfourthprecentTopthirdSEActivityNum}{12.99\%\xspace}

\newcommand{\RQfourthtopfourthSEActivity}{\emph{software security}\xspace}
\newcommand{\RQfourthprecentTopfourthSEActivityNum}{10.39\%\xspace}
\newcommand{\RQfifthtopfirstSEProblemNum}{9\xspace}
\newcommand{\RQfifthtopfirstSEProblem}{\emph{scheduling problem}\xspace}

\newcommand{\RQfifthtopsecondSEProblem}{\emph{test suite minimization}\xspace}

\newcommand{\RQfifthtopthirdSEProblemNum}{7\xspace}
\newcommand{\RQfifthtopthirdSEProblem}{\emph{software failure prediction}\xspace}

\newcommand{\RQfifthtopfourthSEProblem}{\emph{join order problem}\xspace}

\newcommand{\RQfifthtopfifthSEProblem}{\emph{security attack identification}\xspace}

\newcommand{\RQfifthtotalNumSEProblem}{32\xspace}
\newcommand{\RQfifthSEProblemAcrossPhasefirst}{\emph{scheduling problem}\xspace}
\newcommand{\RQfifthSEProblemAcrossPhasesecond}{\emph{security attack identification}\xspace}
\newcommand{\RQfifthSEProblemAcrossPhasethird}{\emph{security management}\xspace}
\newcommand{\RQfifthSEProblemAcrossPhasefourth}{\emph{software failure prediction}\xspace}

%% file: title.tex
Quantum Optimization for Software Engineering: A Survey

%% file: abstract.tex
Quantum computing, particularly in the area of quantum optimization, is steadily progressing toward practical applications, supported by an expanding range of hardware platforms and simulators. While Software Engineering (SE) optimization has a strong foundation, which is exemplified by the active Search-Based Software Engineering (SBSE) community and numerous classical optimization methods, the growing complexity of modern software systems and their engineering processes demands innovative solutions. This Systematic Literature Review (SLR) focuses specifically on studying the literature that applies quantum or quantum-inspired algorithms to solve classical SE optimization problems. We examine \finalselected primary studies selected from an initial pool of \totalNumInitalSearch publications obtained through systematic searches of six digital databases using carefully crafted search strings. Our findings reveal concentrated research efforts in areas such as SE operations and software testing, while exposing significant gaps across other SE activities. Additionally, the SLR uncovers relevant works published outside traditional SE venues, underscoring the necessity of this comprehensive review. Overall, our study provides a broad overview of the research landscape, empowering the SBSE community to leverage quantum advancements in addressing next-generation SE challenges.

%% file: keywords.tex
Quantum Optimization, Quantum-inspired Algorithm, Classical Software Engineering, Systematic Literature Review

%% file: content.tex

\input{intro}

\section{Background}
\label{sec:background}

\input{background}


\input{researchmethod}


\input{empiricalstudy}

\section{Threats To Validity}
\label{sec:threats}
First, our SLR coverage spans from 2014 to 2025 and is limited to six major digital libraries. While this may not capture every publication, the relatively recent emergence of quantum and quantum-inspired applications in software engineering means that this ten-year window likely encompasses the vast majority of relevant work. We believe this time frame adequately reflects the field’s development and current state.

Second, we followed \swebok to define and scope SE activities and problems. Although alternative frameworks could lead to different classifications, \swebok is widely recognized and comprehensive, providing a solid foundation for our study. Regarding study selection and labelling, the initial screening was conducted by one author based on carefully discussed search strings agreed upon by all co-authors. All authors participated in the labelling and agreed upon the results. This process was designed to minimize bias and ensure rigour. 

\section{Related Work}
\label{sec:relatedwork}

\begin{table}[!th]
    \small
    \centering
    \caption{Summarization of related review articles within QSE. Note that the header \textit{\# of Primary Studies} corresponds to the number of papers collected for the literature review, instead of all the involved references. The content \textit{Unspec.} indicates the review paper does not specify this item.}
    \label{tab:rw-qse}
    \resizebox{.98\textwidth}{!}{
        \input{tables/related-qse}
    }
\end{table}

\begin{table}[!th]
    \small
    \centering
    \caption{Summarization of related review articles for quantum-related optimization algorithms.}
    \label{tab:rw-opt}
    \resizebox{.98\textwidth}{!}{
        \input{tables/related-opt}
    }
\end{table}
\input{related}

\section{Conclusion}
\label{sec:conclusions}

This Systematic Literature Review (SLR) is the first to comprehensively examine the literature on applying quantum and quantum-inspired algorithms to classical software engineering (SE) optimization problems. It provides a thorough overview of the current state of the art in this emerging interdisciplinary area. Given the cross-disciplinary nature of the field, such an overview is essential to bridge gaps between different research communities—an importance underscored by the fact that many primary studies identified in the review were published outside traditional SE venues. Through this SLR, we highlight significant opportunities for leveraging quantum and quantum-inspired algorithms to better address classical SE optimization challenges. 
Our objective with this review is to inspire related communities to recognize these opportunities and fosters further research and collaboration to advance next-generation SE solutions.

\section*{Acknowledgments}
This work is supported by the State Key Laboratory of Complex \& Critical Software Environment (SKLCCSE, grant No. CCSE-2024ZX-01). We acknowledge the contribution of Peiru Li to the initial search process of the systematic literature review.

%% file: intro.tex
\section{Introduction}
Based on Richard Feynman's insight in ``Simulating Physics with Computers''~\cite{feynman1982simulating}, quantum computing seeks to efficiently simulate nature by leveraging computers whose operations obey the principles of quantum mechanics (e.g., superposition and entanglement) to avoid the limits of classical computers in simulating the quantum behaviour of physical systems.
By leveraging quantum bits (qubits), which can exist in multiple states simultaneously, quantum computers promise to solve certain problems exponentially faster than classical computers across various domains~\cite{farhi2000quantum, kitaev2002classical}. 
In particular, quantum optimization devices such as quantum annealers are designed to solve combinatorial optimization problems by mapping them onto energy minimization tasks, offering practical applications even in the current \textbf{Noisy Intermediate-Scale Quantum (NISQ)} era~\cite{yarkoni2022quantum, quinton2025quantum}. 

Classically, numerous algorithms and heuristics such as \textbf{Simulated Annealing (SA)}, \textbf{Genetic Algorithms (GA)}, and \textbf{Non-dominated Sorting Genetic Algorithm II (NSGA-II)} have been proposed to address various optimization problems~\cite{delahaye2018simulated_annealing, lambora2019genetic-algorithms, yusoff2011nsga-ii}. These classical approaches often achieve (near-)optimal solutions within practically acceptable time frames, even for large-scale and complicated \textbf{Software Engineering (SE)} activities, like software testing~\cite{cohen2003augmenting}, automatic program repair~\cite{le2011genprog}, and software refactoring~\cite{mkaouer2014high}. However, with the emergence and advances of specialized quantum hardware (e.g., quantum annealers such as D-Wave and gate-based quantum computers such as IBM Quantum and Google Sycamore), researchers have increasingly focused on exploring whether quantum optimization algorithms can outperform their classical counterparts in solving a diverse set of optimization problems. This shift aims to determine if quantum methods offer advantages in computational speed, solution quality, or scalability for complex optimization challenges. For instance, D-Wave has been applied for vehicle routing optimization~\cite{irie2019quantum} and test optimization~\cite{wang2024test}, and gate-based quantum hardware like IBM Quantum has been used for portfolio optimization~\cite{buonaiuto2023best} and job scheduling~\cite{wang2024qoncord}.
The emergence of these applications is also due to the availability of quantum optimization algorithms such as Grover Search~\cite{grover1996fast} and \textbf{Quantum Approximate Optimization Algorithm (QAOA)}~\cite{moll2018quantum}, which are well-suited for exploring complex solution spaces during optimization~\cite{abbas2024challenges}.  

Moreover, an active research field explores how the principles of quantum mechanics can be integrated into classical optimization algorithms such as \textbf{Particle Swarm Optimization (PSO)}~\cite{alvarez2021three}, which are termed as quantum-inspired algorithms and aim to bring the claimed performance advantages without needing quantum computers. This can be considered as a pragmatic bridge between established optimization techniques and emerging quantum computing paradigms before scalable quantum hardware becomes widely accessible. For instance, \textbf{Quantum-Inspired Evolutionary Algorithms (QIEAs)}~\cite{montiel2019quantum} have been proposed to extend classical evolutionary algorithms (e.g., GAs) to explore the solution space more effectively and use binary encoding akin to qubits to cope with SE problems like requirement selection~\cite{\papertwentyeighth} and test suite generation~\cite{\papersixtysixth}. 

Though the field of \textbf{Quantum Software Engineering (QSE)} is still in its early stages, the community is very active in proposing solutions that apply best practices of classical SE for engineering quantum software~\cite{TOSEMVision2025, TOSEMRoadmap2025}. The QSE community also investigated how to leverage quantum optimization to benefit classical SE practices. For instance, Wang et al.~\cite{wang2024test} proposed BootQA for test case minimization and Ammermann et al.~\cite{ammermann2024quantum} employed QAOA for configurations of product lines. However, applications of quantum(-inspired) optimization for classical SE still remain underdeveloped. There remains a critical gap in comprehensive overviews of available solutions, established best practices within the QSE community and beyond, and attempted applications
in this emerging interdisciplinary domain. 

To this end, in this paper, we present the first \textbf{Systematic Literature Review (SLR)} of the literature involving quantum and quantum-inspired optimization for classical SE. Our SLR provides a deep analysis of 77 primary studies published between 2014 and 2025, and answers 11 \textbf{Research Questions (RQs)}. Particularly, we zoom into the publication trends, the addressed SE activities and problems, the proposed solutions, and the identified challenges. In the last RQ, we also outline the open research challenges based on the findings from empirical studies, which could offer guidelines for future research on applying quantum-based solutions to SE problems. For replicability and reproducability, we have the artifact of our SLR available at \hyperlink{https://github.com/WSE-Lab/QuantumOpt4SE}{https://github.com/WSE-Lab/QuantumOpt4SE}.

The paper structure: Section~\ref{sec:background} presents the background and preliminary knowledge related to quantum and quantum-inspired optimization. Section~\ref{sec:researchmethod} introduces the RQs concerned in our paper and elaborates on the methodology adopted to collect primary studies. Results and analyses for answering each RQ are reported in Sections~\ref{sec:publication}-\ref{sec:openChallenges}, followed by threats to validity in  Section~\ref{sec:threats}. In Section~\ref{sec:relatedwork}, we present the related work and then we conclude the paper in Section~\ref{sec:conclusions}.

%% file: background.tex
In this section, we discuss quantum computing, QSE, and quantum and quantum-inspired optimization, which are required background for smoothly comprehending the rest of the paper. 

\subsection{Quantum Computing}
From Jim Baggott’s historical narrative in the quantum story~\cite{baggott2011quantum}, exploring the conceptual foundations of quantum mechanics, to the contemporary realization of quantum computing across diverse hardware platforms, we witness the rapid evolution of quantum technologies and their growing applications in domains such as cryptography~\cite{gisin2002quantum}, optimization~\cite{abbas2024challenges}, and finance~\cite{herman2023quantum}, over years.  

Particularly, when looking at the history of quantum computing, its start is signified by Paul Benioff’s proposal of the quantum Turing machine~\cite{baggott2011quantum} in the 1980s, marking the convergence of quantum mechanics and computer science. Later on, in 1995, Juan Ignacio Cirac and Peter Zoller proposed the Cirac–Zoller controlled-NOT gate~\cite{cirac1995quantum}, an implementation of the controlled-NOT (CNOT) quantum logic gate, which built the foundation for ion-trap quantum computing. After this proposal, quantum computing hardware developed rapidly across various platforms, including trapped ions (e.g., IonQ\footnote{IonQ: \hyperlink{https://ionq.com/}{https://ionq.com/}}) and superconducting qubits (e.g., IBM's quantum computers\footnote{IBM's quantum computers: \hyperlink{https://www.ibm.com/quantum}{https://www.ibm.com/quantum}}). In 2011, D-Wave\footnote{D-Wave: \hyperlink{https://www.dwavequantum.com/}{https://www.dwavequantum.com/}} introduced \textbf{Quantum Annealing (QA)} hardware specialized for optimization tasks. Nowadays, ion traps, superconducting circuits, photonics, and neutral atoms are all being actively developed, which together drive toward scalable and fault-tolerant quantum computers. Along with the development of quantum hardware, quantum programming languages have been proposed, such as Q\# from Microsoft, Qiskit from IBM, Cirq from Google, and many others. 

When we zoom into quantum optimization hardware that leverages actual quantum mechanical effects such as superposition, entanglement, and quantum tunnelling, there are mainly two categories: (1) quantum annealers from D-Wave, quantum devices that exploit quantum tunnelling and superposition to find low-energy states of Ising-like models; and (2) gate-based quantum computers from IBM, Google, IonQ, and others, which use universal quantum gates to implement algorithms like \textbf{Quantum Approximate Optimization Algorithm} (QAOA) for optimization.

Spanning the current landscape, there also exists hardware that does not rely on actual quantum mechanics but is inspired by quantum models and behaviours. For instance, \textbf{Coherent Ising Machines (CIMs)} (e.g., the 100,000 spin CIM developed by NTT Japan~\cite{honjo2021100}) are considered as analog devices because they use optical components such as lasers to emulate quantum behaviours and exploit coherence and interference of optical waves to find low-energy states. Meanwhile, digital quantum-inspired solvers like Fujitsu’s \textbf{Digital Annealer (DA)}\footnote{Fujitsu's DA: \hyperlink{https://www.fujitsu.com/global/services/business-services/digital-annealer/}{https://www.fujitsu.com/global/services/business-services/digital-annealer/}} use classical digital hardware to efficiently tackle complex combinatorial optimization problems by simulating QA processes. 

Quantum computer simulators (e.g., IBM’s Qiskit Aer, Google’s Cirq) are classical software that emulates the behaviour of quantum computers by mathematically representing qubits and quantum gates on classical computers. These simulators are very useful for enabling researchers to design and test quantum algorithms without having access to real quantum hardware. In the QSE community, compared to real quantum hardware, quantum simulators are available and affordable backends for running quantum programs. Therefore, simulators are commonly applied to ensure the quality of quantum software in the view of code, via techniques including software testing~\cite{li2025preparation} and debugging~\cite{huang2019statistical}, program repair~\cite{li2024automatic}, and so on. 

\subsection{Quantum and Quantum-inspired Optimization}
Optimization is commonly considered a process of finding the optimal (best) solution from a set of feasible ones, subject to specified constraints. It involves iteratively improving a system, design, or decision to achieve one or more objectives, e.g., minimizing cost, maximizing performance, or balancing tradeoffs within given limitations. 
In mathematical terms, an optimization problem (usually stated in terms of minimization) can be defined as: 
\[
\begin{aligned}
\text{Minimize} \quad & f(\mathbf{x}) \\
\text{subject to} \quad & g_i(\mathbf{x}) \leq 0, \quad i = 1, 2, \ldots, m, \\
& h_j(\mathbf{x}) = 0, \quad j = 1, 2, \ldots, p, \\
& \mathbf{x} = [x_1, x_2, \dots, x_n]
\end{aligned}
\]
where $f(\mathbf{x})$ is the objective function; $\mathbf{x}$ is a vector of decision variables: $\mathbf{x} = [x_1, x_2, \dots, x_n]$; $h_j(\mathbf{x})$ are equality constraints; and $g_i(\mathbf{x})$ are inequality constraints. 

In classical SE, optimization involves improving various aspects of the \textbf{Software development Life Cycle (SDLC)} of software systems, such as enhancing code performance~\cite{petke2017genetic, yuan2018arja}, refining software design~\cite{aleti2012software, filieri2015automated}, optimizing resource allocation~\cite{ngo2008optimized, wang2010multi}, and improving testing processes through techniques like test suite minimization~\cite{tallam2005concept, wang2015cost}, test case prioritization~\cite{elbaum2002test, pradhan2019employing}, and automated test generation~\cite{ali2009systematic, jia2015learning}.
Over the years, the \textbf{Search-Based Software Engineering (SBSE)} community\footnote{The SBSE community: https://ssbse.com/} has emerged as an active and influential venue for applying meta-heuristic and evolutionary optimization techniques, e.g.,
\textbf{Genetic Algorithm (GA)},
\textbf{Simulated Annealing (SA)},
and \textbf{Particle Swarm Optimization (PSO)},
to address various SE optimization problems.

Quantum optimization includes solving optimization problems using quantum hardware (or corresponding classical simulators), which includes approaches like QA and gate-based algorithms like QAOA on universal quantum computers. Quantum-inspired optimization does not use quantum hardware but draws inspiration from quantum principles. Quantum-inspired algorithms are executed on classical hardware and fall into two categories: (1) specialized hardware (e.g., Fujitsu’s DA), where algorithms must be tailored to the system’s architecture; and (2) general-purpose classical hardware, where extensions of known algorithms (e.g., GA, PSO) integrate quantum-inspired concepts for enhanced performance. 
We summarize the key differences among quantum optimization, quantum-inspired optimization with specialized hardware, and quantum-inspired optimization with general hardware in Table~\ref{tab:ComparingOptimizationMethods}.

\begin{table}[!t]
	\small
	\centering
	\caption{Quantum Optimization Paradigms}
	\label{tab:ComparingOptimizationMethods}
    \resizebox{.98\textwidth}{!}{
        \input{tables/background.tex}
    }
    \begin{spacing}{0.8}
    	\raggedright \footnotesize 
    	DA: Digital Annealer;
    	CIM: Coherent Ising Machine;
    	CPU: Central Processing Unit;
    	GPU: Graphics Processing Unit;
    	FPGA: Field-Programmable Gate Array;
    	DA: Digital Annealing;
    	NTT: Nippon Telegraph and Telephone Corporation;
    	QAOA: Quantum Approximate Optimization Algorithm;
    	QA: Quantum Annealing;
    	QUBO: Quadratic Unconstrained Binary Optimization;
    	GA: Genetic Algorithm;
    	PSO: Particle Swarm Optimization.
    \end{spacing}
\end{table}

\subsection{Quantum Algorithms for Optimization}
To bridge the gap between SE researchers with quantum computing, we introduce several applicable quantum algorithms that can serve to optimize a search procedure or solve an optimization problem.

\subsubsection{Preliminaries}
Below, we provide preliminaries related to quantum information, and more details can be found in \cite{nielsen2010quantum}. In classical computing, bits are the basic computational unit that only represents binary value, i.e., either 0 or 1. By contrast, qubits in place of bits serve for quantum computing and quantum superposition enables qubits to exist in a state other than $\ket{0}$ and $\ket{1}$~\footnote{The ket $\ket{\;}$ is one of Dirac notations in quantum mechanics.}. Formally, a pure single-qubit quantum state can be denoted by the linear combination of its computational basis states $\ket{0}$ and $\ket{1}$: 
\begin{equation}
    \label{eq: single-qubit-state}
    \ket{\psi}=\alpha \ket{0} + \beta \ket{1}, 
\end{equation}
where the complex numbers $\alpha$ and $\beta$ are amplitudes of the two states and subject to $|\alpha|^2 + |\beta|^2 = 1$. Furthermore, quantum entanglement facilitates interactions between qubits, allowing them to share information in a way that more complicated quantum states can be represented and manipulated, which, however, cannot be achieved with unentangled qubits. 

The time evolution of the quantum state of a closed quantum system is described by the Schr\"{o}dinger equation,
\begin{equation}
    i \hbar \frac{\text{d}\ket{\psi(t)}}{\text{d}t}=\hat{H}\ket{\psi(t)},
\end{equation}
where $\ket{\psi(t)}$ is the state of the quantum system at the time $t$; $\hat{H}$ refers to an observable called Hamiltonian; $\hbar$ is the Planck's constant; and $i$ is the imaginary unit ($i=\sqrt{-1}$), a non-real complex number. QA relies on continuous-time evolution under the guidance of the adiabatic theorem. Qubits start in a superposition of all candidate states with equal amplitudes and evolve according to governed by the time-dependent Schr\"{o}dinger equation. In contrast, gate-based computers introduce a generic quantum circuit model to depict the evolution of qubits. In this model, qubit states evolve through a sequence of discrete-time steps. For example, the state transformation from step $t$ to step $t+\Delta t$ can be formulated by $\ket{\psi(t+\Delta t)}=U\ket{\psi(t)}$. Operator $U$ is unitary and called a quantum (logic) gate. Quantum gates are the building blocks of quantum circuits, and their performance is analogous to classical logic gates for conventional digital circuits. 
There are many basic quantum gates used in gate-based quantum computers, and they can be represented by unitary matrices, such as the Pauli-X gate $X$ (a.k.a., NOT gate) and the Hadamard gate $H$ for operation on a single qubit, and the controlled-NOT gate $CNOT$ involving two qubits:
$$
X \coloneqq \left[\begin{array}{cc}
     0 & 1\\
     1 & 0 \\
    \end{array}\right],
\qquad
H \coloneqq\frac{1}{\sqrt{2}}\left[\begin{array}{cc}
     1 & 1\\
     1 & -1 \\
    \end{array}\right], 
\qquad
CNOT \coloneqq\left[\begin{array}{cccc}
     1 & 0 & 0 & 0 \\
     0 & 1 & 0 & 0 \\
     0 & 0 & 0 & 1 \\
     0 & 0 & 1 & 0 \\
    \end{array}\right].
$$

Quantum measurement is the only way for an observer in the classical world to extract information from a closed quantum system. Especially, the observed outcome is not deterministic but probabilistic. 
According to the Born rule, the probability of obtaining a specific outcome is associated with the amplitude of the basis state into which this outcome is encoded. In detail, once state $\ket{\psi}$ defined by Equation~(\ref{eq: single-qubit-state}) is measured, the probabilities of obtaining $0$ and $1$ are $|\alpha|^2$ and $|\beta|^2$, respectively. However, upon quantum measurement, this state will inevitably collapse to one of the basis states, i.e., either $\ket{0}$ or $\ket{1}$.

\subsubsection{Grover Search}
Grover Search~\cite{grover1996fast} is a well-known quantum search algorithm giving theoretical support for quantum advantage. This algorithm is specifically designed for a search problem of querying an unstructured database with size $N$, where the querying can be denoted as a mapping
$f:\{0,1,\cdots, N-1\}\rightarrow\{0,1\}$ and then this problem is formalized as finding one or multiple target indices $x$ satisfying $f(x)=1$ 
with high probability. Grover Search can demonstrate quadratic speedup over classical computing: from $O(N)$ to $O(\sqrt{N})$. 

Grover Search is implemented on gate-based quantum computers. The algorithm includes three components and each search iteration executes the oracle and diffusion operations: (1) \textit{state initialization}: a superposition state is prepared by acting Hadamard gates on the default state, and the computational basis states composing the superposition refer to possible indices of the database; (2) \textit{oracle operation}: the oracle operation on the initialized state indicates the querying action, and then the indices are simultaneously split into a target group and non-target one; and (3) \textit{diffusion operation}: due to the computational basis states of the two groups assigned with different amplitudes, the diffusion plays a role in amplifying the amplitude corresponding to the target group, such that the intended indices can be observed with larger probability. After a pre-defined number of iterations, the measurement outcome is interpreted as the index $x$.
Looking back on the search process, quantum superposition enables the parallel distinction of the target indices from others with a single query. In contrast, the classical brute-force approach can only determine whether a single index is the target or not, requiring sequential checks of candidate indices.

\subsubsection{Quantum Annealing}
QA~\cite{apolloni1989quantum, apolloni1990numerical} is an optimization procedure based on the adiabatic theorem for finding the global minimum of a given objective function over a set of candidate solutions. Compared to the widely used heuristic optimization algorithm, i.e., simulated annealing, which takes inspiration from the physical process of annealing and is implemented classically, QA physically utilizes quantum phenomena like superposition, entanglement, and quantum tunneling, enabling the state to escape from local minima and reach global minima more efficiently.  

When solving a combinatorial optimization problem, a minimization objective function should be defined in a mathematical expression that can represent the Hamiltonian (i.e., the sum of kinetic and potential energy) of the quantum system. This expression is converted into models with binary variables, like the Ising or \textbf{Quadratic Unconstrained Binary Optimization (QUBO)} model, for the implementation on the quantum hardware. The QA system then maps the binary variables to be solved onto physical qubits. During the physical procedure of annealing, the qubits involved in the quantum system are prepared in a superposition state at first. This superposition indicates the combination of all possible states, fulfilling possible quantum parallelism. Next, the system evolves, guided by quantum dynamics, toward the lowest valley in the energy landscape, which is denoted as the optimal solution to the combinatorial optimization problem. This evolution depends on a parameter known as the annealing time, which determines the rate of change in the system's energy and plays a crucial role in ensuring the intended convergence to the global minimum.

Nevertheless, the physical qubits in the NISQ annealer are not fully connected, meaning that each physical qubit can entangle with only a few others. As one solution, D-Wave relies on a qubit chain to group multiple physically connected qubits to logically represent one binary variable, which, however, sacrifices the total number of available logical physical qubits. Such quantum devices require extremely low temperatures to maintain coherence and minimize environmental interference. 
Instead, Fujitsu proposed DA, a specialized classical computing architecture, which solves combinatorial optimization problems by digitally emulating some principles of QA, including quantum tunnelling and parallelism using room-temperature digital circuits. 

\subsubsection{Variational Quantum Algorithms (VQA)}
VQAs~\cite{cerezo2021variational} are a family of hybrid quantum-classical algorithms that optimize parameterized quantum circuits (ans\"{a}tze) by iteratively minimizing a cost function 
(i.e., objective function). This optimization is performed through classical updates of circuit parameters based on measurement outcomes from \textbf{Quantum Processing Unit (QPU)}, i.e., a state-of-the-art processing hardware that uses qubits to solve complex problems using quantum mechanics.
QAOA and \textbf{Variational Quantum Eigensolver (VQE)} follow the hybrid computational paradigm of VQAs.

In QAOA~\cite{farhi2000quantum}, the objective function can be formulated as Ising or QUBO, and then encoded as a cost Hamiltonian $H_C$ in QAOA. QAOA initializes all qubits in an equal superposition state by applying a Hadamard gate to each qubit, creating a uniform probability distribution over all possible computational basis states. This also implies that QAOA's initial state is agnostic to the problem. After the initialization, the optimization procedure has three main steps: (1) \textit{cost operation}: unitary operator $U_C(\gamma_j)=\exp(-i\gamma_j H_C)$ is associated with cost Hamiltonian $H_C$, a variational parameter $\gamma_j$ is appended to the quantum circuit, where $j$ refers to the operation at the $j$-th layer 
as users can configure the number of layers $p$ (such that $j=1,2,\cdots,p$), which determine the repetitive executions for the cost and the following mixing Hamiltonians before quantum measurement; (2) \textit{mixing operation}: another unitary operation $U_M(\beta_j)=\exp(-i\beta_j H_M)$ following $U_C(\gamma_j)$, where $H_M$ is the mixing Hamiltonian that enables QAOA to explore different parts of the search space and avoid getting stuck in local optima; (3) \textit{ansatz updating}: after $p$ repetitions of the cost and mixing operations, the final quantum state is measured, and then the classical optimizer (e.g., gradient descent) solves the optimal parameters $\boldsymbol{\gamma}\coloneqq(\gamma_1,\cdots,\gamma_p), \boldsymbol{\beta}\coloneqq(\beta_1,\cdots,\beta_p)$, which results in the minimum or maximum expectation value of $H_C$ and updates the raw variational parameters in the ansatz. QAOA iterates with the above three steps until the stop condition (e.g., a pre-defined maximum number of iterations) is satisfied.

Unlike QAOA, which is primarily designed for combinatorial optimization with problem-agnostic ans\"{a}tze, VQE ~\cite{peruzzo2014variational}  targets the ground-state estimation of physical systems.  While QAOA relies on a cost Hamiltonian derived from a classical objective function, VQE directly implements \textit{problem-specific} ans\"{a}tze composed of controlled gates and rotation gates, where the rotation angles correspond to variational parameters. This structure allows VQE to natively encode quantum mechanical operators, such as molecular Hamiltonians, without requiring combinatorial problem mapping. 


%% file: tables/background.tex
\begin{tabular}{l   l   p{0.35\textwidth}   p{0.35\textwidth}}
    \toprule 
    Aspect & Quantum Optimization & Quantum-inspired Optimization with Specialized Hardware & Quantum-inspired Optimization with General Hardware \\
    \midrule
    Hardware & Quantum processors & Classical and custom-designed (e.g., DA, CIM) & Standard classical (CPU, GPU, FPGA) \\
    \midrule
    Example platforms & D-Wave, IBM Q & Fujitsu DA, NTT CIM & Any classical device \\
    \midrule
    Algorithm type & Quantum-native (QAOA, QA) & QUBO/Ising-based, tailored & Quantum-inspired variants of GA, PSO, etc. \\
    \bottomrule
\end{tabular}

%% file: researchmethod.tex
\section{Research Method}
\label{sec:researchmethod}

In this section, we discuss the research method we applied for conducting this SLR, including RQs, digital libraries and search strings, paper selection criteria, snowballing, and data extraction.  

\subsection{Research Questions}
\label{subsec:rq}

To gain a comprehensive view on how SE problems are tackled with quantum and quantum-inspired optimization, we carried out the SLR to answer the following \totalNumRQs RQs from four aspects:

\begin{itemize}
	\item Bibliometric Analysis
	\begin{itemize}
		\item \rqAYear
		\item \rqAVenue
		\item \rqAPublicationType
	\end{itemize}
	\item Addressed SE Activities and Problems
	\begin{itemize}
		\item \rqBSEActivity
		\item \rqBSEProblem
		\item \rqBSEReformulation
	\end{itemize}
	\item Proposed Quantum and Quantum-inspired Solutions
	\begin{itemize}
		\item \rqCQOType
		\item \rqCQOInspired
		\item \rqCQOComputing
	\end{itemize}
	\item Open Challenges
	\begin{itemize}
		\item \rqEClaimed
		\item \rqEOpen
	\end{itemize}
\end{itemize}

\subsection{Databases and Search Queries}
\label{subsec:queries}
To be comprehensive, we collected all published literature that is relevant to answer our RQs through searching six digital databases: 
\dbACM, \dbIEEE, \dbScopus, \dbSpringer, \dbWoS, and \dbWiley (alphabetical order).
These databases were selected based on the observation that they were frequently used in SLR
studies recently published in the SE venues of high quality, such as~\cite{zhou2024large, li2025systematic}. 

To investigate recent advances in quantum and quantum-inspired optimization techniques for solving SE problems, we define five sets of search terms as shown in Listing~\ref{lst:searchquery}.
The terms ``quantum'' (Set~\ref{set:quantum}) and ``software'' (Set~\ref{set:software}) are defined to reduce the scope of the search to publications, such that they are relevant to quantum and software only. 
The terms in 
Set~\ref{set:op} indicates that we are only interested in the literature that is relevant to optimization. 
The terms in Set~\ref{set:technique}
covers techniques used for optimization, while
the terms in Set~\ref{set:se} 
collectively covers core SE phases or activities 
referred to another recent SLR~\cite{liu2024large} that presents a detailed and comprehensive analysis of applications for SDLC.
To accommodate the different database search engines,
the search queries eventually used to conduct this SLR 
are summarized in Table~\ref{tab:queries}.

\begin{lstlisting}[style=mystyle, escapechar=©,caption={Search Queries}, label={lst:searchquery}]
quantum AND # Set 1 ©\label{set:quantum}©
(optimi* OR minimi* OR prioriti* OR selection OR reduction) AND # Set 2: optimization problem ©\label{set:op}©
(algorithm OR heuristic* OR search OR learning OR artificial intelligence OR AI) AND # Set 3: technique ©\label{set:technique}©
(api OR develop* OR bug OR code OR coding OR debug OR defect OR deploy OR evolution OR fault OR fix OR maintenance OR program OR refactor* OR repair OR requirement OR test OR verification OR validation OR vulnerab* OR configur*) AND # Set 4: software engineering phases/activities ©\label{set:se}©
software # Set 5 ©\label{set:software}©
\end{lstlisting}

\subsection{Paper Selection Criteria}
\label{subsec:selection}
\begin{figure}[!t]
    \centering
    \includegraphics[width=0.98\linewidth]{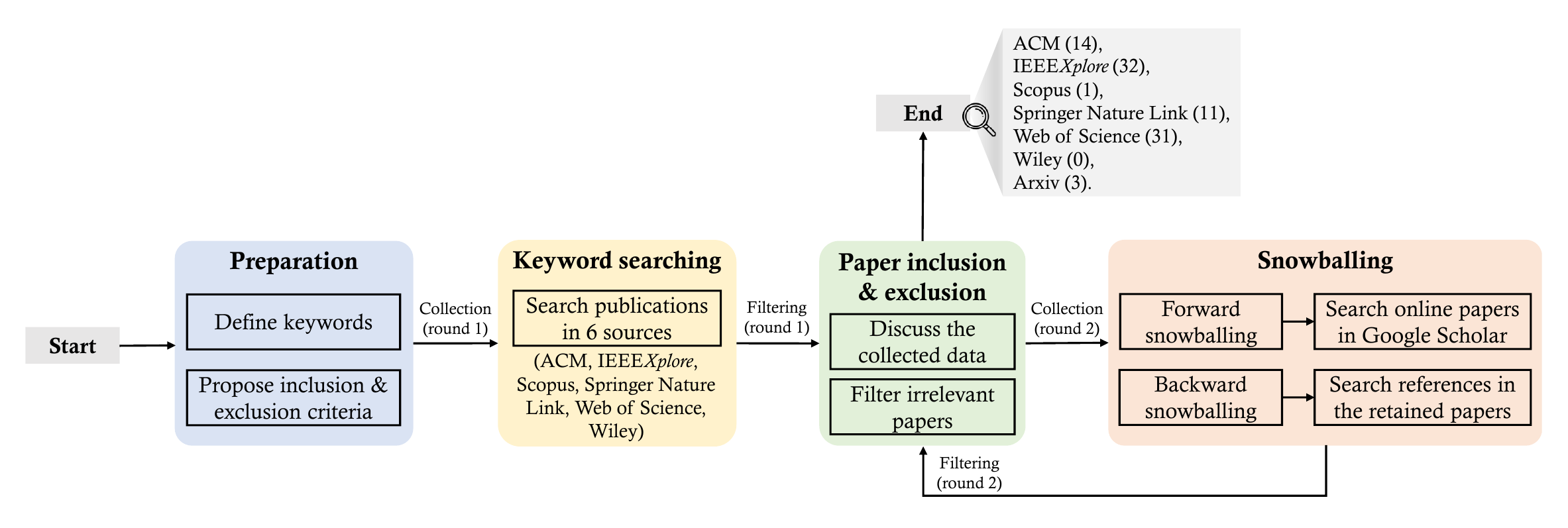}
    \caption{The framework of the adopted methodology for collecting and filtering primary studies}
    \label{fig:method-diagram}
\end{figure}

The framework of the adopted methodology is illustrated in Fig.~\ref{fig:method-diagram}, where we marked the number of finally collected papers for each source, and one paper could be obtained via multiple sources.
As shown in Table~\ref{tab:queries}, we conducted the initial search using the defined search queries across six databases on \initialSearchDate, yielding \initialtotal papers published between 2014 and 2025.
To ensure the relevance of the selected studies, we defined inclusion criteria outlined in Table~\ref{tab:inclusionCriteria} to further guide the paper selection process.
Only papers written in English were considered. 
In addition, each paper was required to address at least one specific SE problem, ensuring the application context. 
To align with the focus of our study, the papers also needed to be related to optimization problems and either employ or propose quantum-based solutions.
We excluded theses, and secondary studies such as surveys or other SLRs.
Secondary studies related to this SLR are discussed in Section~\ref{sec:relatedwork} (Related Work).

The selection was performed by all authors, and the final decision to include each paper was also confirmed by all authors.
By following the criteria, we selected \finalquery papers.

\begin{table}
	\small
	\centering
	\caption{Inclusion and Exclusion Criteria}
	\label{tab:inclusionCriteria}
	\input{tables/inclusion-criteria.tex}
\end{table}

\subsection{Snowballing}
\label{subsec:snowballing}

We also applied the snowballing strategy to avoid potential missing publications. 
From the~\finalquery selected studies, 
on~\snowballingSearchDate, we systematically performed backward and forward snowballing as many iterations as necessary until no more publications were further identified. During the process, we also applied the paper selection criteria defined in Table~\ref{tab:inclusionCriteria}. Three authors of the paper discussed intensively during the snowballing until a consensus was reached. Eventually from the snowballing process, we obtained~\finalsnowballing publications. Together with the~\finalquery publications already selected, we, in total, obtained~\finalselected publications as the primary studies of our study.

\subsection{Data Extraction}
\label{subsec:extraction}

During data extraction, we created an initial spreadsheet, which includes metadata, such as the title, publication year, and venue, for all \finalselected selected papers.
To guide the extraction of additional data, we defined the information required to answer each RQ, as outlined in Table~\ref{tab:data-extraction}, and expanded the spreadsheet with additional columns accordingly.
RQs~1–3 could be addressed using metadata obtained directly from the databases.
However, for RQs~4–10, we conducted a detailed review of each paper to extract the necessary information. 
For example, to answer RQ4, we added a column namely \textit{SE Activity} to capture specific SE activities discussed in each paper. 
The terminologies used for these activities are based on the definitions from 
Software Engineering Body of Knowledge v4.0~\cite{swebok2024} (abbreviated as \swebok for convenience).
For RQ6, we collected both the number of objectives being optimized in the context of the SE problems and the number of objectives reformulated in optimization and then categorized each problem as \textit{single}, \textit{multiple}, or \textit{many}-objective optimization based on widely used definitions in SBSE~\cite{li2015many, ramirez2018systematic}. In detail, (1) single-objective optimization problems are defined in terms of a unique objective; (2) multi-objective optimization problems are characterized by the presence of 2 or 3 objectives that usually conflict with each other; and (3) many-objective optimization problems particularly refer to requiring the definition of 4 or more objectives
In addition, we identified whether there are constraints reformulated.
By studying papers during the data extraction, the information required to be extracted can be refined.
For instance, when investigating the development and running environment for solutions with quantum computing (RQ9), we expanded our data collection to include how classical approaches were integrated into these solutions.

By following Table~\ref{tab:data-extraction}, we went through each paper to collect the relevant information.
All extracted data for each paper can be found at \hyperlink{https://github.com/WSE-Lab/QuantumOpt4SE}{https://github.com/WSE-Lab/QuantumOpt4SE}.

\begin{table}
	\small
	\centering
	\caption{Information to be collected corresponding to each research question}
	\label{tab:data-extraction}
		\input{tables/data-extraction.tex}
\end{table}

%% file: tables/inclusion-criteria.tex
\begin{tabular} { l l }\\ 
	\toprule 
	\# &  Inclusion and Exclusion Criterion \\ 
	\midrule 
	1 & The paper is written in English \\ 
	2 & The paper must target specific SE problem(s). \\ 
	3 & The paper must relate to optimization problem(s).\\
	4 & The paper must either employ or propose quantum related solutions. \\ 
	5 & The paper is not a thesis. \\
	6 & The paper is not a survey or a systematic literature review. \\ 
	\bottomrule 
\end{tabular} 

%% file: tables/data-extraction.tex
\begin{tabular}{ l  p{.8\textwidth}}
	\toprule 	
	RQs     & Information being collected \\
	\midrule
	RQ1		& Number of papers published per year \\ 
	RQ2		& Number of papers published per venue \\	
	RQ3		& 1) Publication type per paper \\
			& 2) Number of the papers by type \\	
	RQ4		& 1) Type(s) of SE activities targeted in each paper \\
			& 2) Number of papers associated with each SE activity \\
	RQ5		& 1) SE problem(s) tackled in each paper \\
			& 2) Number of papers associated with each SE problem \\
	RQ6		& 1) Number of objectives optimized for each SE problem in each paper \\ 
			& 2) Number of reformulated objectives for each SE problem in each paper \\
			& 3) Whether the proposed solution has constraints\\
			& 4) Number of papers addressing single- vs. multiple-objectives in optimization reformulation \\
	RQ7		& 1) Types of quantum optimization (i.e., classical, quantum computing, hybrid) per paper \\
			& 2) Number of papers associated with each type \\
	RQ8		& 1) Whether the proposed solution is quantum-inspired for each paper classified as classical or hybrid\\
			& 2) Algorithm extended/adopt and principles applied in each quantum-inspired solution \\
			& 3) Number of papers associated with each algorithm and each applied principle \\
	RQ9		& 1) Quantum algorithm applied (e.g., QA) used for each paper classified as quantum computing or hybrid, along with their categories (e.g., mathematical model) and development/running environment (e.g., backend, programming toolkit) \\
			& 2) Number of papers associated with each category and each type \\
	RQ10 	& Research challenges identified by each paper \\
	RQ11	& Open research challenges remain unaddressed \\
	\bottomrule 
\end{tabular}

%% file: empiricalstudy.tex
\section{Bibliometric Analysis}
\label{sec:publication}

In this section, we study the trends of publications for quantum and quantum-inspired optimization for classical SE in terms of time (RQ1), venue (RQ2), and venue type (RQ3).

\subsection{\rqAYear}

\begin{figure}
	\centering
	\includegraphics[width=.6\textwidth]{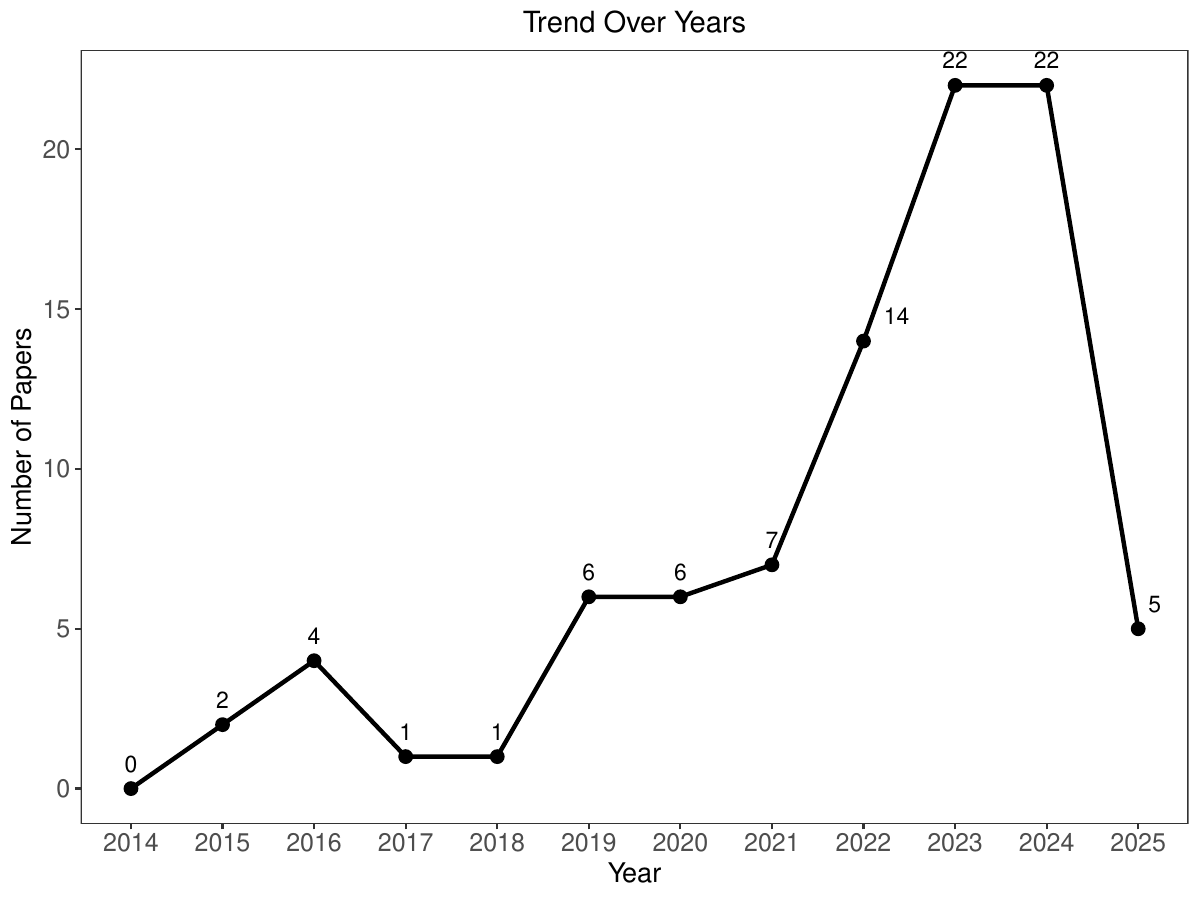}
	\caption{Number of publications over the years (2014-2025)}
	\label{fig:rq1_plot_year}
\end{figure}

To identify trends in publications over the years, Figure~\ref{fig:rq1_plot_year} illustrates the number of papers published each year from 2014 to 2025. Despite 2014 being within the temporal range of keyword searching, we did not find intended papers published in that year. From the figure, we can see that the field began gaining more attention starting in 2019, followed by a sharp increase in publications from 2021 onward, peaking at \RQfirstmaxNumYear papers in both \RQfirstmaxYearfirst and \RQfirstmaxYearsecond. Note that our database search concluded on \initialSearchDate, and snowballing ended on \snowballingSearchDate, such that only 5 papers newly available in 2025 are included in our SLR. The observed upward trend suggests the field will maintain its momentum, offering opportunities to tackle more optimization challenges through novel quantum and quantum-inspired algorithms, with increasingly accessible and reliable quantum hardware.

\begin{results}[Findings for RQ1]
	Since 2019, quantum and quantum-inspired optimizations have gained increasing attention for tackling classical SE problems.
\end{results}

\subsection{\rqAVenue}

Table~\ref{tab:rq2_venue} presents a list of publication venues along with the number of papers published in each venue. 
Venues with only a single publication were grouped under \textit{Other} with their venue type, e.g., \textit{Other Journal}.
Based on Table~\ref{tab:rq2_venue}, it can be observed that the \finalselected primary studies have been published in a very diverse set of venues. Specifically, \RQsecondmaxNumConference primary studies were published in the \RQsecondmaxConference and all the rest were all published in different conference venues. This indicates that the field is highly interdisciplinary. Moreover, the \RQsecondmaxConference demonstrates its potential to establish itself as a core venue/community of the field. For journal publications, for instance, 3 primary studies were published in \RQsecondmaxJournalfirst; 2 and 2 primary studies were published in the top SE venues: TSE and TOSEM, respectively. On the other hand, there are 29 primary studies published in 29 different journal venues. It is clear that there is no central venue for the topic. These results indicate that although the topic is about quantum and quantum-inspired optimization for classical SE, none of the SE venues is the primary publication target, implying that the topic is an emerging topic and highly interdisciplinary, and there is a need for new forums to bridge different communities.

\begin{table}
	\small
	\centering
	\caption{Number of publications per venue}
	\input{generated_files/RQ2_venue.tex}
	\label{tab:rq2_venue}
\end{table}

\begin{results}[Findings of RQ2]
	Research on quantum and quantum-inspired optimization for classical SE problems has been published across a broad range of venues, indicating that the topic is both emerging and highly interdisciplinary. 
\end{results}
\subsection{\rqAPublicationType}

\begin{figure}
	\centering
	\includegraphics[width=.6\textwidth]{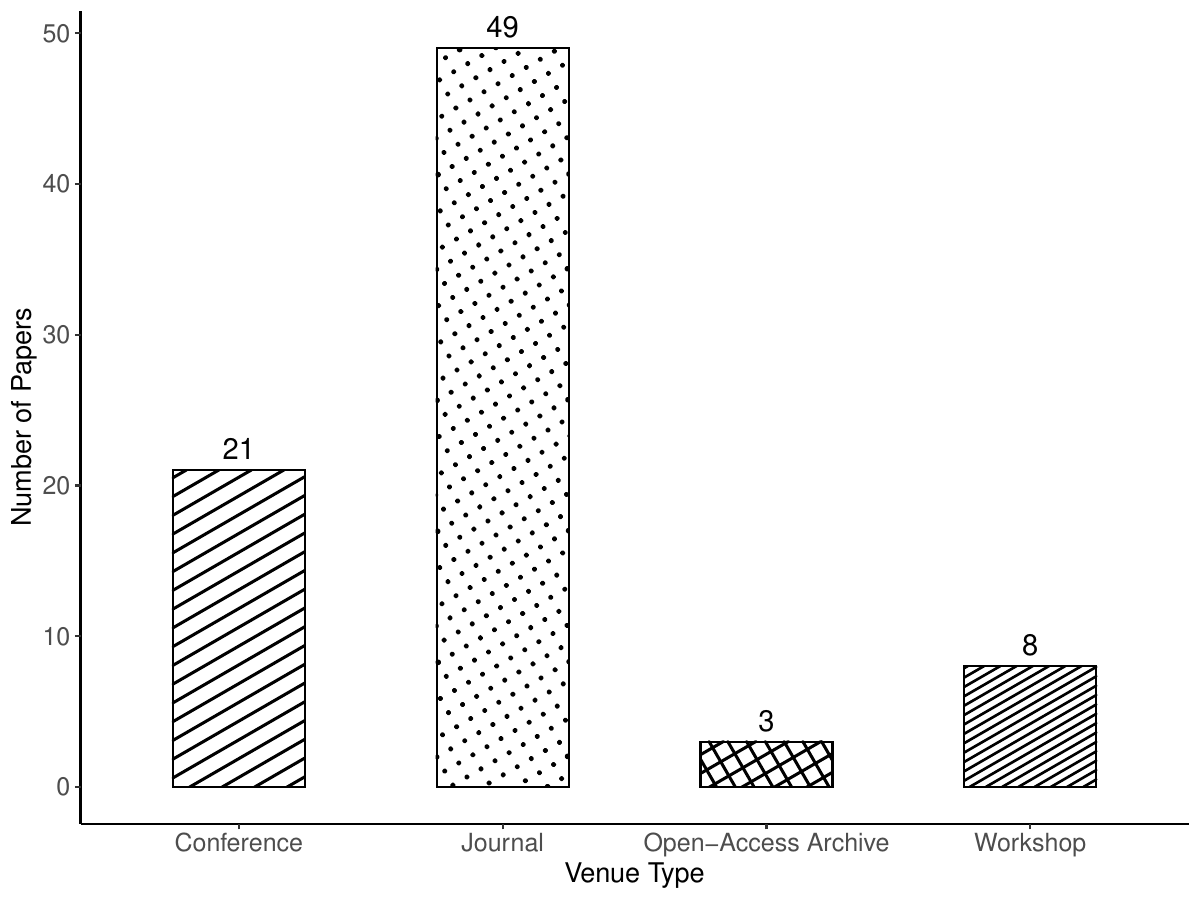}
	\caption{Number of publications per venue type}
	\label{fig:rq3_publication_type}
\end{figure}

Analyzing publication types, to a certain extent, can provide us insights into research maturity and dissemination trends, we classify the publication types into four categories and present them in Figure~\ref{fig:rq3_publication_type}. As shown in the figure, journal publications dominate the primary studies (\RQthirdtotalNumJournal papers, \RQthirdprecentJournal), significantly outnumbering conference papers (\RQthirdtotalNumConference papers, \RQthirdprecentConference), workshop publications (\RQthirdtotalNumWorkshop papers, \RQthirdprecentWorkshop), and open-access archive entries (\RQthirdtotalNumOpenAccessArchive papers, \RQthirdprecentOpenAccessArchive). This distribution suggests that the majority of included primary studies offer validated knowledge, as journal publications typically are associated with a more rigorous peer-review process. 

\begin{figure}
	\centering
	\includegraphics[width=.5\textwidth]{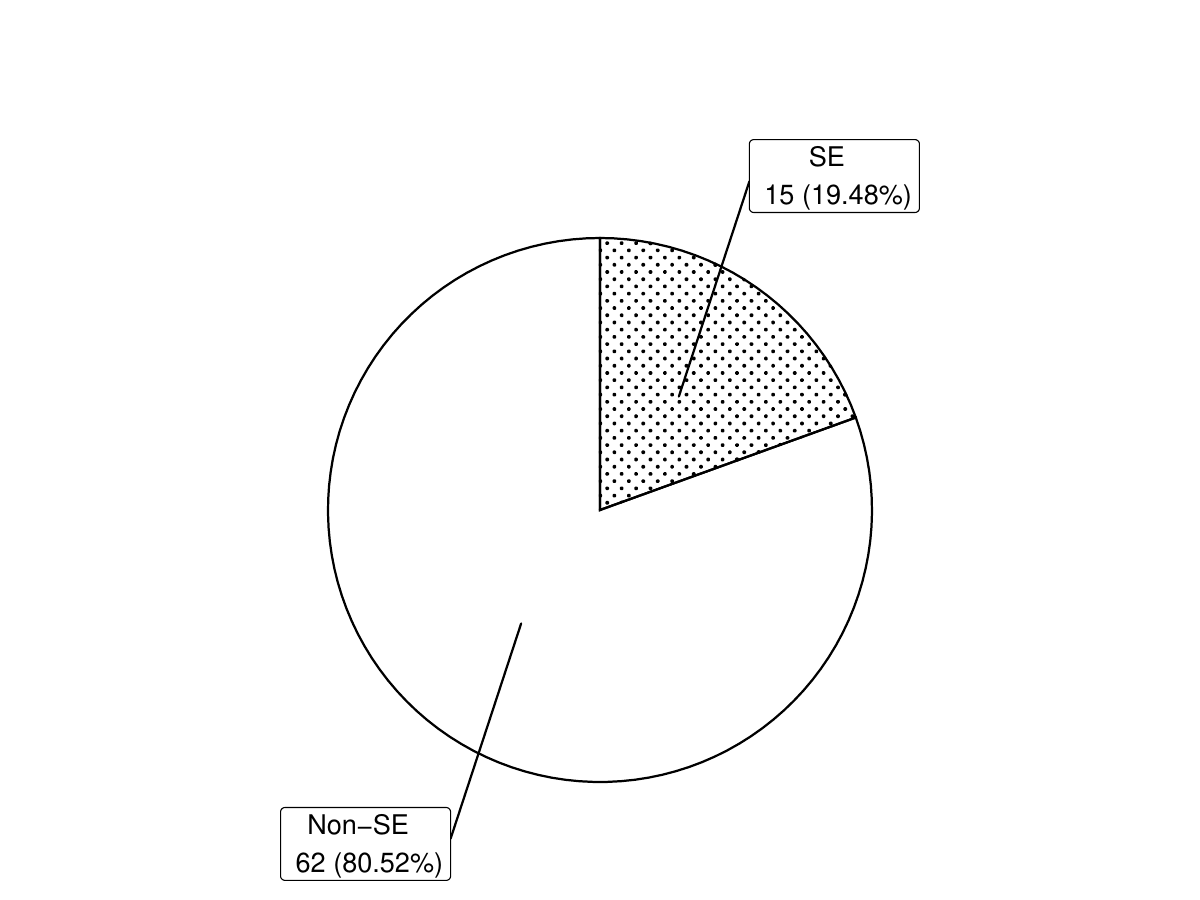}
	\caption{Number of primary studies belongs to SE or Non-SE}
	\label{fig:rq2_venue_se}
\end{figure}

As shown in Figure~\ref{fig:rq2_venue_se}, we observed that among the \finalselected primary studies, only \RQthirdNumSEVenue of them were published in the SE venues, while the majority (\RQthirdprecentNonSEVenue) appeared in other disciplinary venues (e.g., \RQsecondmaxJournalsecond). This distribution suggests that there is a significant interdisciplinary engagement with the topic beyond the SE research community. This distribution also indicates that our SE community should also underscore the importance for SE researchers to look into relevant work published in other disciplines, as significant contributions to this domain are emerging beyond SE venues. Moreover, this observation shows that it is indeed necessary to conduct this SLR to gain a comprehensive view of the literature.

\begin{results}[Findings of RQ3]
	Among the \finalselected primary studies, \RQthirdprecentJournal have been published in journals, surpassing \RQthirdprecentConference in conferences, \RQthirdprecentWorkshop in workshops, and \RQthirdtotalNumOpenAccessArchive in open-access archive entries.
	This suggests that the majority of the primary studies are solid research outcomes and exhibit a growing maturity and academic interest in the field.
	However, among the \finalselected primary studies, only \RQthirdprecentSEVenue of the studies have been published in SE venues, highlighting the need for new interdisciplinary forums to better connect the SE community with advances in quantum and quantum-inspired optimization. 
\end{results}

\section{Addressed SE Activities and Problems}
\label{sec:se}

In this section, we analyze the selected primary studies from the SE perspective: the SE activities addressed (RQ4), the specific SE problems tackled (RQ5), and the types of optimization problems they were reformulated into (RQ6), i.e., single-, multiple-, or many-objective problems.

\subsection{\rqBSEActivity}\label{subsec:SEactivity}

\begin{figure}
	\centering
	\includegraphics[width=.8\textwidth]{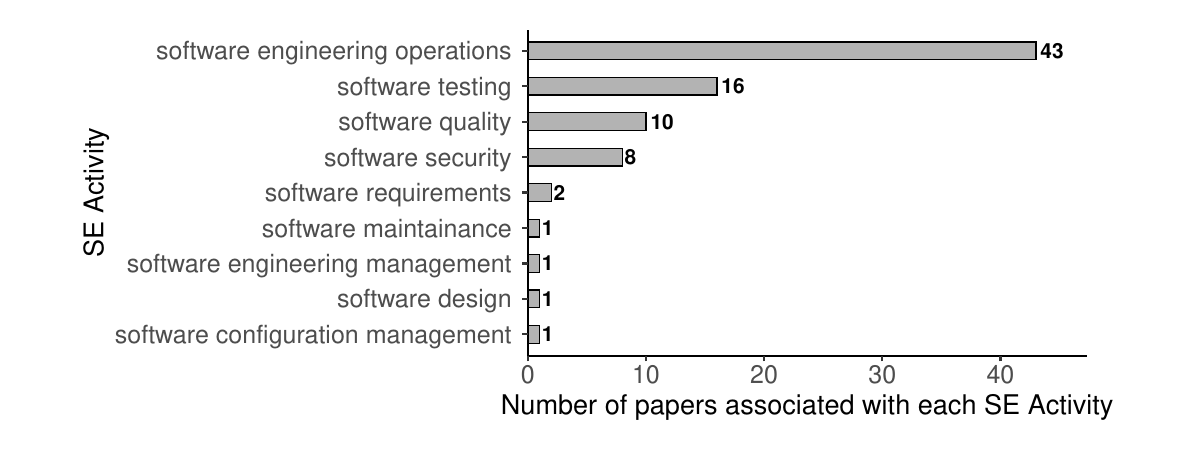}
	\caption{Bar plot representing the distribution of papers across SE activities}
	\label{fig:rq4_se_activity}
\end{figure}

Considering that this SLR is about studying quantum and quantum-inspired optimization in classical SE, we look into specific SE activities the primary studies target. To be systematic, we referred to the widely recognized SE body of knowledge~\cite{swebok2024} to obtain the classification of SE activities. 
The number of studies targeting each SE activity is presented in Figure~\ref{fig:rq4_se_activity}.

From Figure~\ref{fig:rq4_se_activity}, we can observe that \RQfourthprecentTopfirstSEActivityNum of primary studies (i.e., \RQfourthtopfirstSEActivityNum papers) fall under the \RQfourthtopfirstSEActivity category, defined as \emph{``the set of activities and tasks necessary to deploy, operate and support a software application or system while preserving its integrity and stability''} (SWEBOK~\cite{swebok2024}). The remaining studies address \RQfourthtopsecondSEActivity (\RQfourthprecentTopsecondSEActivityNum), \RQfourthtopthirdSEActivity (\RQfourthprecentTopthirdSEActivityNum), and \RQfourthtopfourthSEActivity (\RQfourthprecentTopfourthSEActivityNum). Other SE activities are significantly under-represented, with only 1-2 studies each, suggesting these areas remain largely unexplored in the existing literature. From the perspective of SDLC, the primary studies cover most SDLC phases, including requirements, design, configuration, testing, operation and maintenance, quality, and management. 

\begin{results}[Findings of RQ4]
	The study of the literature exhibits a strong research focus on the \RQfourthtopfirstSEActivity activities (\RQfourthprecentTopfirstSEActivityNum of the \finalselected primary studies) of SDLC.
	As these activities often face non-deterministic, complex, and dynamic problems, quantum and quantum-inspired optimization techniques might offer promising potential for cost-effective solutions.
	The remaining studies mostly focus on software quality assurance–related activities, i.e., \RQfourthtopsecondSEActivity (\RQfourthprecentTopsecondSEActivityNum), \RQfourthtopthirdSEActivity (\RQfourthprecentTopthirdSEActivityNum), and \RQfourthtopfourthSEActivity (\RQfourthprecentTopfourthSEActivityNum)
\end{results}

\subsection{\rqBSEProblem}

\begin{figure}
	\centering
	\includegraphics[width=.8\textwidth]{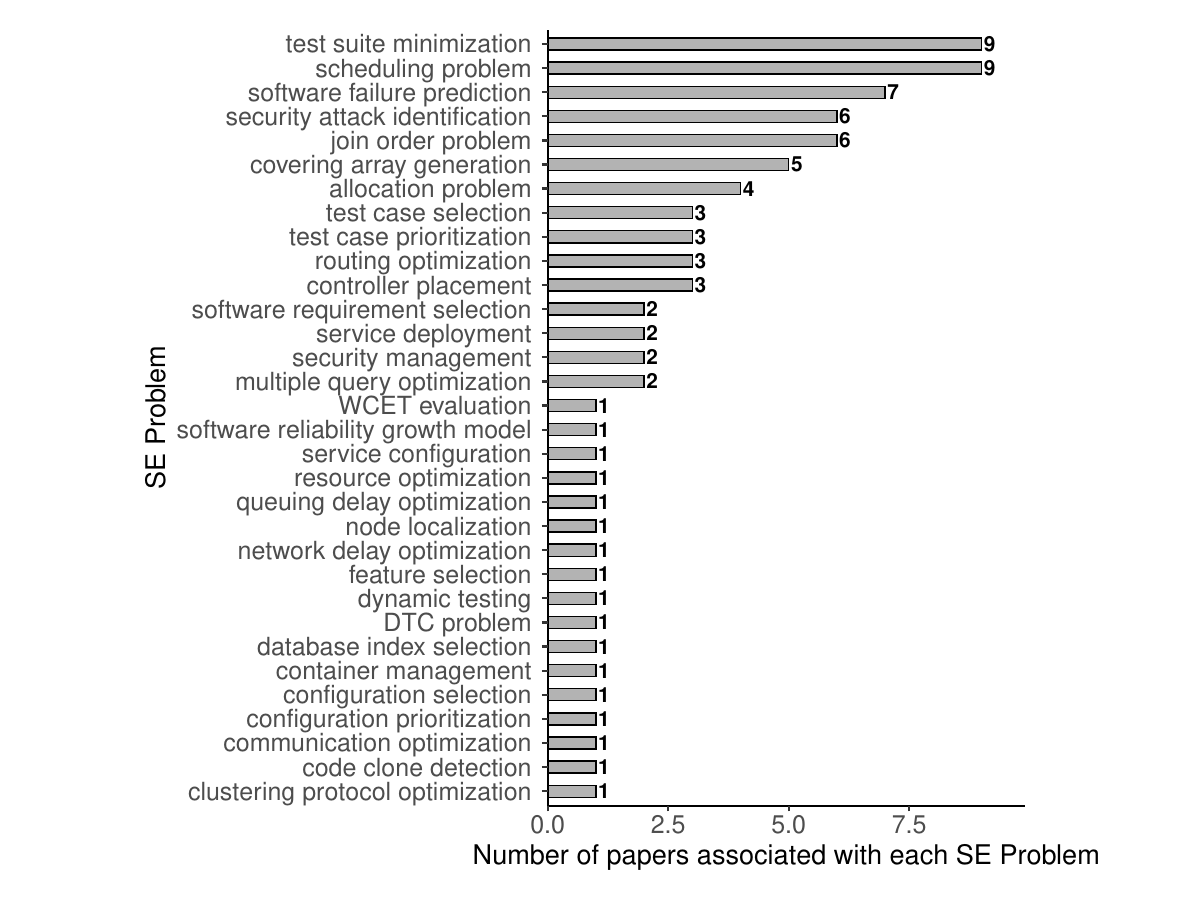}
	\caption{Bar plot representing the distribution of papers for SE problem}
	\label{fig:rq5_se_problem}
\end{figure}

When further looking into specific SE problems addressed by the primary studies, from Figure~\ref{fig:rq5_se_problem}, one can observe that the 
\finalselected 
primary studies, in total, target 
\RQfifthtotalNumSEProblem 
different problems. This suggests a broad and diverse research landscape. Furthermore, we can observe that 
\RQfifthtopfirstSEProblem 
and 
\RQfifthtopsecondSEProblem 
were targeted the most (with 
\RQfifthtopfirstSEProblemNum
primary studies for each problem), followed by 
\RQfifthtopthirdSEProblem (with 
\RQfifthtopthirdSEProblemNum 
primary studies). This observation shows that while some well-established SE problems (e.g., test suite optimization, scheduling, failure prediction) dominate the literature, many other issues remain understudied. In addition, this observation well aligns with the broader trends in classical SE research, where testing-related topics (e.g., test suite minimization, test case selection, test case prioritization, software failure prediction) have historically been among the most extensively studied areas, as software testing is costly within the SDLC. An existing finding from questionnaire-based statistics supports that over half of projects spend over 20\% of the time testing their systems, and even more than 15\% of projects do testing with more than 40\% of their time~\cite{zhao2003quality}.

In Table~\ref{tab:rq5_se_problem}, we further show the alignment between the SE problems and the corresponding SE activities (from SWEBOK).  
From the table, we can observe that problems belonging to \RQfourthtopfirstSEActivity are very diverse (i.e., 19 different problems), including scheduling of tasks and resources, join order and multiple query optimization in databases, service deployment and configuration, resource optimization, and node localization. The \RQfourthtopsecondSEActivity problems cover different testing phases: test optimization (i.e., test suite minimization, test case prioritization, test case selection), test generation (i.e., covering array generation in combinatorial testing), and test execution (i.e., dynamic testing). The \RQfourthtopthirdSEActivity category covers several important problems: software failure prediction, dynamic transitive closure (DTC) problem in program analysis, reliability evaluation, and worst case execution time (WCET) analysis. 
In \RQfourthtopfourthSEActivity, 6 out of 8 approaches address \RQfifthtopfifthSEProblem for various domains, such as automotive software~\cite{\papersixth}, Internet of Things (IoT) systems~\cite{\paperfiftyfifth,\papersixtyninth}, networks~\cite{\papertwentysecond}, big data platforms~\cite{\paperfortysecond}.

\begin{table}
	\small
	\centering
	\caption{Papers for tackling SE problems}
	\label{tab:rq5_se_problem}
	\resizebox{.9\textwidth}{!}{
		\input{generated_files/RQ5_problem.tex}
	}
\end{table}

As shown in Table~\ref{tab:rq5_se_problem}, there exists four SE problems related to multiple SE activities (denoted with various colous of ~$\blacktriangleright$), i.e., \RQfifthSEProblemAcrossPhasefirst, \RQfifthSEProblemAcrossPhasesecond, \RQfifthSEProblemAcrossPhasethird, and \RQfifthSEProblemAcrossPhasefourth.
Regarding the \RQfifthSEProblemAcrossPhasefirst category (see \textcolor{green}{$\blacktriangleright$} in Table~\ref{tab:rq5_se_problem}), the problems related to tasks performed during operation are categorized under \RQfourthtopfirstSEActivity, such as job scheduling~\cite{\papertwentyfourth}, task scheduling~\cite{\paperfortyfirst}, load scheduling~\cite{\paperfourteenth,\paperninetyfirst}, workflow scheduling~\cite{\papereightysecond}, transaction scheduling~\cite{\paperseventh}, and virtual machine scheduling~\cite{\paperthird}.
Those tasks are mostly addressed in the context of cloud and distributed systems.
In a different context, the problem of project scheduling for enhancing task efficiency~\cite{\paperfortyeighth} is considered part of \textit{software engineering management}.
Regarding \RQfifthSEProblemAcrossPhasesecond (see \textcolor{blue}{$\blacktriangleright$} in Table~\ref{tab:rq5_se_problem}), all problems are associated with \textit{software security}. 
Among them, those identifications performed at runtime are additionally considered part of \RQfifthSEProblemAcrossPhasefirst, i.e., intrusion detection for network security~\cite{\papertwentysecond}, and attack detection for IoT systems~\cite{\paperfiftyfifth,\papersixtyninth}. 
Regarding \RQfifthSEProblemAcrossPhasethird (see \textcolor{orange}{$\blacktriangleright$} in Table~\ref{tab:rq5_se_problem}), two studies~\cite{\papertwentyfifth,\papersixtyeighth} focus on handling security incidents within information security management systems. 
As such, they are relevant to both \RQfourthtopfirstSEActivity and \RQfourthtopfourthSEActivity.
For \RQfifthSEProblemAcrossPhasefourth (see \textcolor{red}{$\blacktriangleright$} in Table~\ref{tab:rq5_se_problem}), all seven studies are categorized under \RQfourthtopfourthSEActivity. Among them, one study~\cite{\papersixtyseventh} also addresses the prediction of cloud service failure time for reliability, and is thus additionally classified under \RQfourthtopfirstSEActivity.
In total, we identified six approaches in software security and quality that extend into the operational phase that might indicating a potential growing trend toward addressing challenges in production environments, exhibiting an increasing convergence of development, deployment, and runtime monitoring to systematically ensure security and quality across the SDLC of software systems.


\begin{results}[Findings of RQ5]
	In total, the \finalselected primary studies cover \RQfifthtotalNumSEProblem distinct SE problems, indicating a broad and diverse research landscape. The top five problems are \RQfifthtopsecondSEProblem, \RQfifthtopfirstSEProblem, \RQfifthtopthirdSEProblem, \RQfifthtopfifthSEProblem and \RQfifthtopfourthSEProblem.
\end{results}

\subsection{\rqBSEReformulation}

\begin{figure}
	\centering
	\includegraphics[width=.6\textwidth]{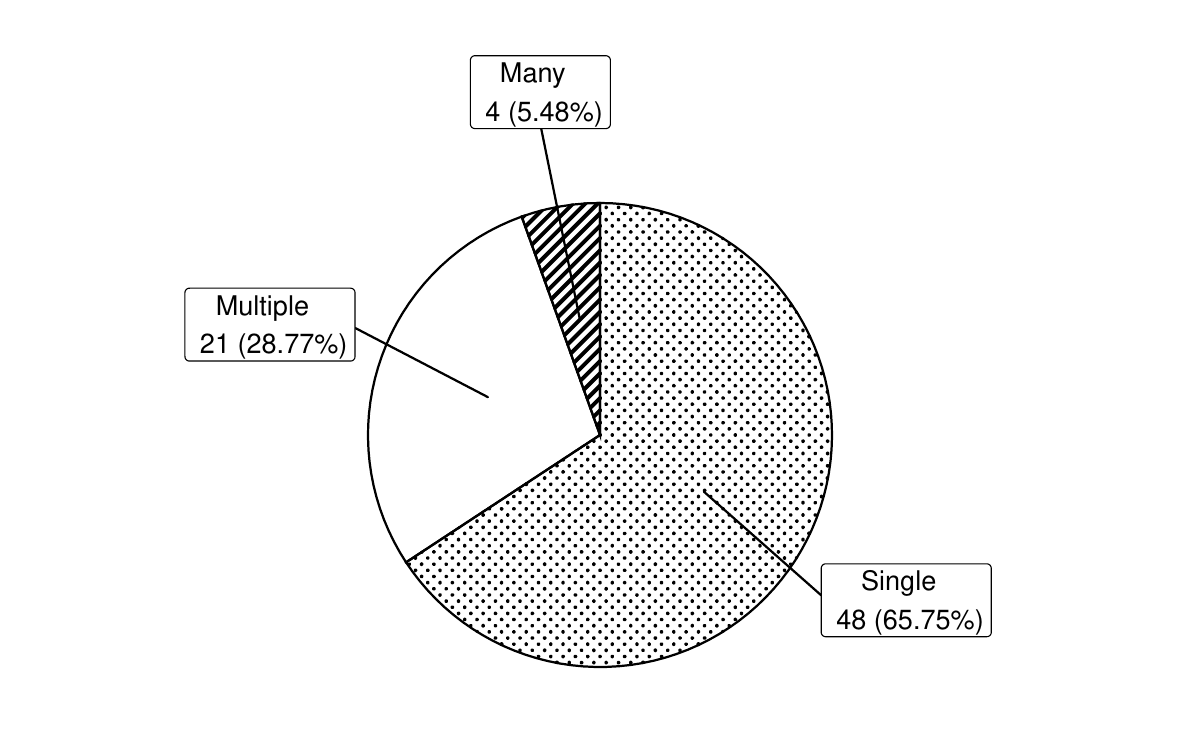}
	\caption{Number of papers associated with \textit{Single}, \textit{Multiple} and \textit{Many} objectives for SE problems}
	\label{fig:rq6_num_prob_objs}
\end{figure}

Optimization problems are typically classified into three categories: single-objective, multiple-objectives and many-objectives. 
Figure~\ref{fig:rq6_num_prob_objs} reports statistics on the types of SE problems addressed in our context, categorized by the number of objectives considered. 
As shown in the figure, among the \finalselected primary studies, 48 papers (65.75\%) address single-objective problems; 21 papers (28.77\%) target multi-objective problems; and only 4 papers (5.48\%) address problems that are many-objectives. From Table~\ref{tab:rq6_reformulation}, we further observe that the four many-objective problems are under \RQfourthtopfirstSEActivity.
In addition, all the four many-objective problems are in the context of IoT and cloud computing, e.g., IoT resource optimization,  virtual machine scheduling in cloud, clustering protocol optimization in energy management, and sensor spacing in IoT networks, mostly due to their inherent need to simultaneously balance four and more conflicting and interdependent objectives to achieve holistic system optimization. 
Regarding multi-objective problems, we also observe that most of approaches (i.e., 15 out of 21) tackle the problems related to  \RQfourthtopfirstSEActivity.
From Table~\ref{tab:rq6_reformulation}, we also observe many single-objective problems such as test suite minimization and software failure prediction. However, many real-world problems in SE have more than one optimization objective, such as 61\% of the collected papers in the SLR~\cite{ramirez2018systematic} with multiple or many-objectives.  
This observation indicates that the simplification has been done to reduce the complexity of real-world problems.  

\begin{table}
	\small
	\centering
	\caption{Reformulation of SE problems in Quantum optimization}
	\label{tab:rq6_reformulation}
	\resizebox{.99\textwidth}{!}{
	\input{generated_files/RQ6_problem_reformulation.tex}}
\end{table}

\begin{figure}
	\centering
	\includegraphics[width=.45\textwidth]{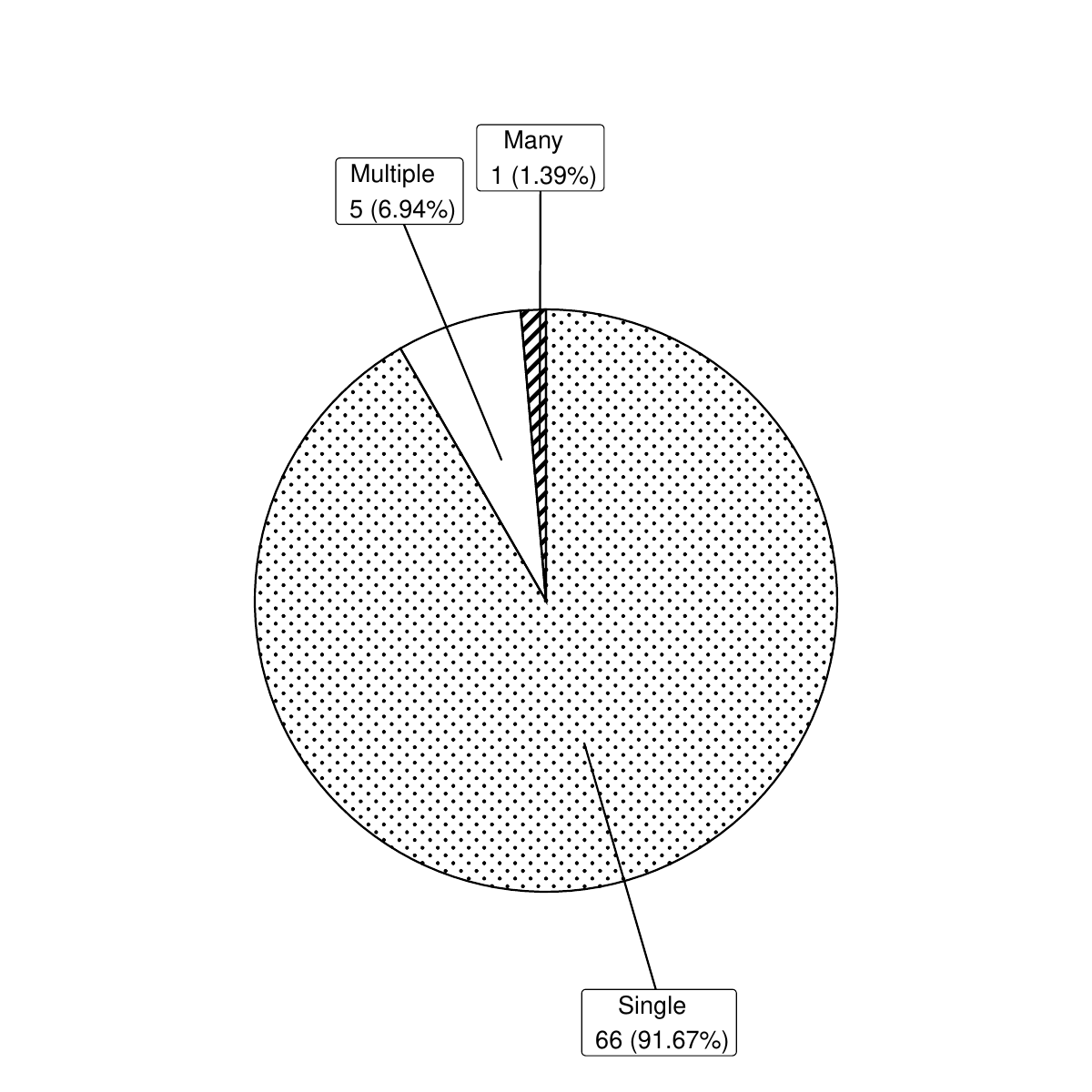}
	\caption{Number of papers associated with \textit{Single}, \textit{Multiple} and \textit{Many} objectives in  reformulation of SE problems}
	\label{fig:rq6_num_objs}
\end{figure}

As shown in Table~\ref{tab:rq6_reformulation}, 
18 out of 25 (72.0\%) many-objective and multi-objective SE problems are reformulated into single-objective formulations.
To solve real-world multi-objective or many-objective problems with quantum or quantum-inspired solutions, these problems are frequently reformulated into single-objective problems primarily due to computational practicality. Quantum algorithms, especially those designed for 
near-term quantum devices (e.g., QAOA), often perform in a simpler manner when optimizing a single cost function, whereas the multiple-objective quantum optimization is required to find a whole set of Parato-optimal solutions via more complex quantum circuits~\cite{ekstrom2025variational}. Combining multiple objectives into a weighted sum for instance and/or relying on constraints to simplify the problem structure make it more compatible with quantum optimization frameworks that operate by minimizing a single energy function, such as an Ising Hamiltonian or QUBO formulation. Moreover, current quantum hardware is constrained by limited number of qubit, short coherence time, imperfect gate fidelities, and 
significant error rates~\cite{preskill2018quantum}. Simplifying the optimization objective helps mitigate these challenges, though doing so may lose the opportunity in capturing trade-offs among competing objectives.
Particularly, when looking at the primary studies of our SLR, as shown in Figure~\ref{fig:rq6_num_objs}, eventually, there are 91.67\% (66 papers) reformulated problems are single-objective, and only 6.94\% (5 papers) and 1.39\% (1 paper) reformulated problems are multi-objective and many-objective. This shows a significant reduction in the number of multi-objective problems, highlighting a strong preference for simplification to enhance practicality in quantum computing.

\begin{figure}
	\centering
	\includegraphics[width=.6\textwidth]{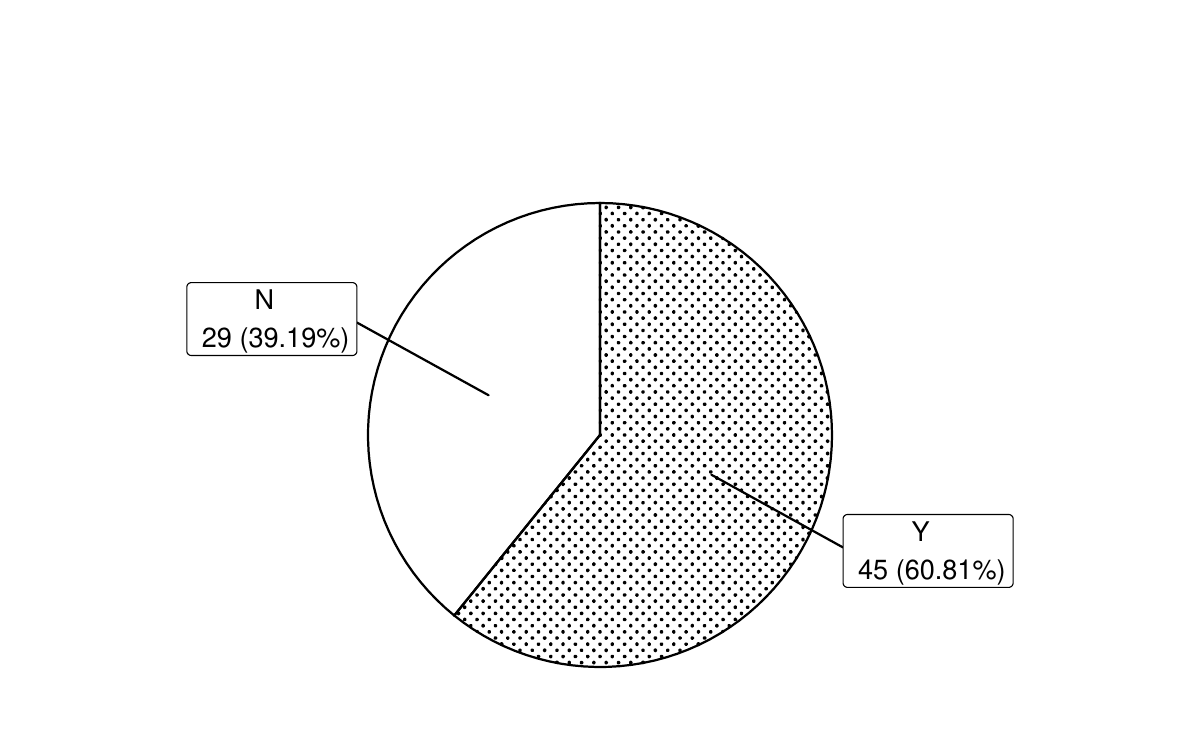}
	\caption{Number of proposed optimizations with defined constraints}
	\label{fig:rq6_constraints}
\end{figure}

In both classical and quantum optimization, constraints are essential in problem formulations, as they guide the search toward valid results.  In classical optimization, constraints are often handled via penalty terms (soft constraints), directly handling in solvers (hard constraints), etc. In quantum or quantum-inspired optimization, handling constraints is one of the biggest challenges. Quantum approaches (e.g., QAOA or QA) typically embed constraints directly into the problem Hamiltonian or cost function, to ensure that low-energy (optimal) quantum states correspond to valid solutions. For instance, same as for classical optimization, constraints can be converted into penalty terms and added to the objective function. For quantum-inspired optimization, constraints are handled similarly to classical metaheuristics such as using penalty terms and constraint-preserving encodings. As shown in Figure~\ref{fig:rq6_constraints}, the formulations of 45 (60.81\%) optimization problems involve constraints. 

\begin{results}[Findings of RQ6]
	Among the \finalselected primary studies, 65.75\% of the SE problems addressed using quantum optimization are single-objective. 
	Most multi-objective and many-objective problems were reformulated into single-objective ones, resulting in 91.67\% of single-objective quantum optimization. Compared to classical SBSE, quantum optimization solutions are less diverse, indicating that quantum optimization for classical SE is still in its early stages, with significant potential for further exploration.
\end{results}


\section{Quantum and Quantum-inspired Solutions}

In this section, we analyze the selected primary studies from three aspect: the types of quantum optimization solutions (RQ7), the inspirations behind quantum-inspired optimization (RQ8), and the application of pure quantum computing algorithms~(RQ9).

\subsection{\rqCQOType}
\begin{figure}
	\centering
	\includegraphics[width=.6\textwidth]{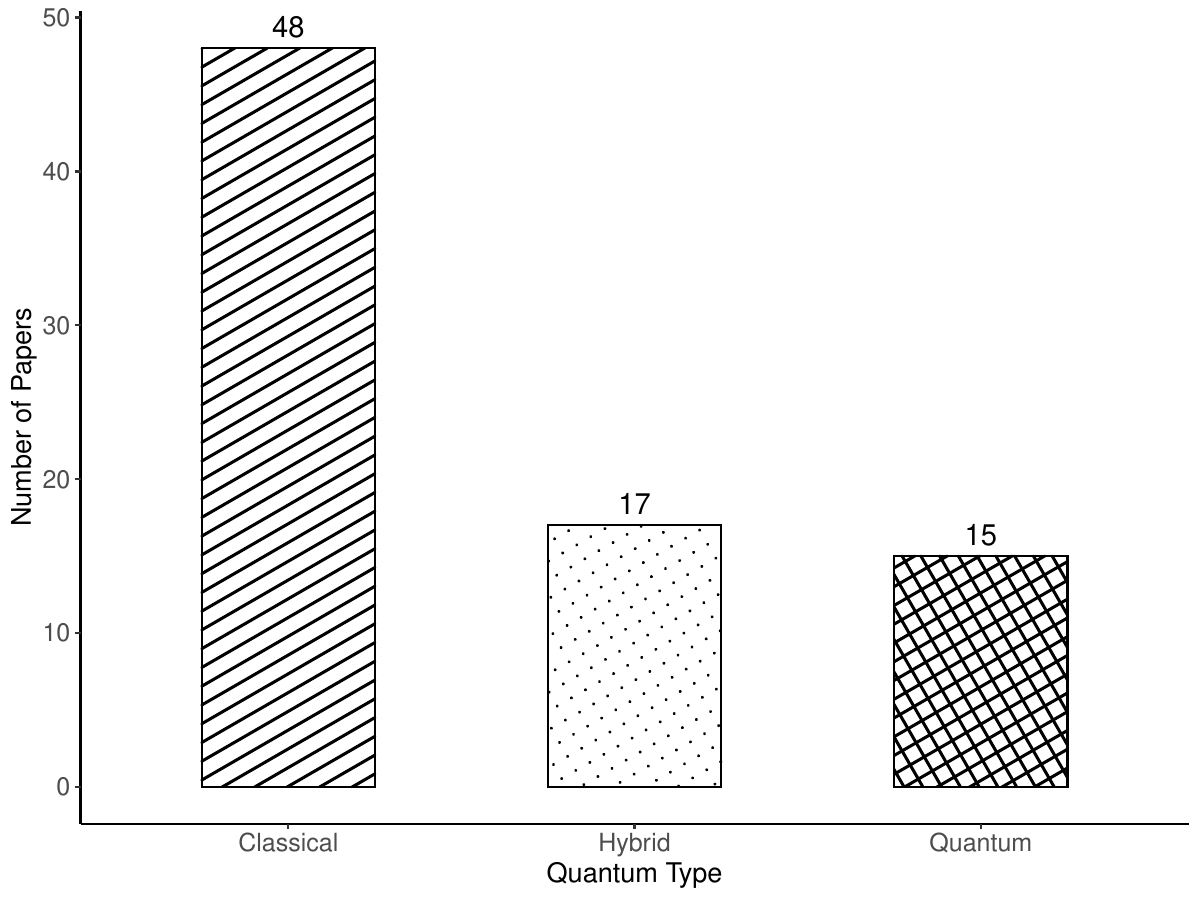}
	\caption{Bar plot representing the distribution of papers associated with each type of quantum or quantum-inspired optimization}
	\label{fig:rq7_qo_type}
\end{figure}

We classify solutions proposed in the primary studies into three categories: classical, hybrid, and quantum, based on the computational resources used for execution. Results are illustrated in Figure~\ref{fig:rq7_qo_type}. From the figure, one can observe that the majority of solutions (48) are classical as they are all quantum-inspired optimization solutions, which are executed on classical computers, while 17 are hybrid as these solutions involve both classical and quantum components. The remaining 15 primary studies propose pure quantum solutions. In Table~\ref{tab:rq7_qo_type}, we list the primary studies of each category for references.

From this observation, we hypothesize that most researchers opted for quantum-inspired optimization algorithms because they are classical and hence can run on existing classical hardware, considering the current quantum hardware has limitations on the scale, noise and accessibility. Quantum-inspired solutions are ideal for exploring if quantum principles can help classical SE problems.
The practical way to utilize limited quantum hardware today for solving large problems is to decompose them into smaller problems that can be solved by available quantum hardware, for instance. In such a case,  often a hybrid solution is proposed. When solving a small problem, it is worth exploring quantum hardware if there is an access.

The proportions are likely to change over time, when quantum hardware becomes more powerful, reliable, and accessible. The proportion of  hybrid and pure quantum solutions are expected to increase significantly, potentially surpassing quantum-inspired. However, quantum-inspired methods are likely to remain valuable and potentially become standard classical optimization techniques as long as evidence demonstrates that they outperform others in solving certain problems.

\begin{table}
	\small
	\centering
	\caption{Papers for each type of quantum or quantum-inspired optimization}
	\label{tab:rq7_qo_type}
		\input{generated_files/RQ7_quantum_type.tex}
	\begin{spacing}{0.8}
		\raggedright \footnotesize
		Note that some papers, e.g., Reference~\cite{bettonte2022quantum}, present more than one approaches categorized under different types, which results in the sum of papers by type is greater than the total number of selected papers.
	\end{spacing}
\end{table}

\begin{results}[Findings of RQ7]
Classical (quantum-inspired) optimization methods currently dominate the literature due to the limitations of existing quantum hardware. However, we anticipate that hybrid and fully quantum optimization solutions will emerge in the future.
\end{results}

\subsection{\rqCQOInspired}
This RQ specifically examines primary studies that propose quantum-inspired optimization solutions. In Table~\ref{tab:rq8_classical}, we summarize the classical algorithms that serve as the foundation for these quantum-inspired approaches, as well as the quantum-relevant mechanisms used to extend or enhance the base algorithms. 


From the table, we can see that PSO is the most frequently used base algorithm, i.e., 16 (33.33\%) quantum-inspired solutions opted for PSO. As a population-based metaheuristic, PSO replies on particles collectively exploring the search space, which conceptually aligns with parallel exploration (e.g., quantum superposition's ability to evaluate multiple states simultaneously). Both conceptually and computationally, PSO is generally
considered simpler than many other search algorithms such as GA including operations of selection, mutation and crossover~\cite{eberhart1998comparison}, which makes PSO a popular choice for quantum-inspired optimization. Though GA is more complex than PSO, according to our study, as shown in Table~\ref{tab:rq8_classical}, it is the second popular choice for quantum-inspired optimization. This might be because GA has a rich set of variants and backed by decades of successful applications in classical optimization~\cite{katoch2021review,tang1996genetic} and hence the research communities are very familiar to GA and its variants.

\begin{figure}
	\centering
	\includegraphics[width=.6\textwidth]{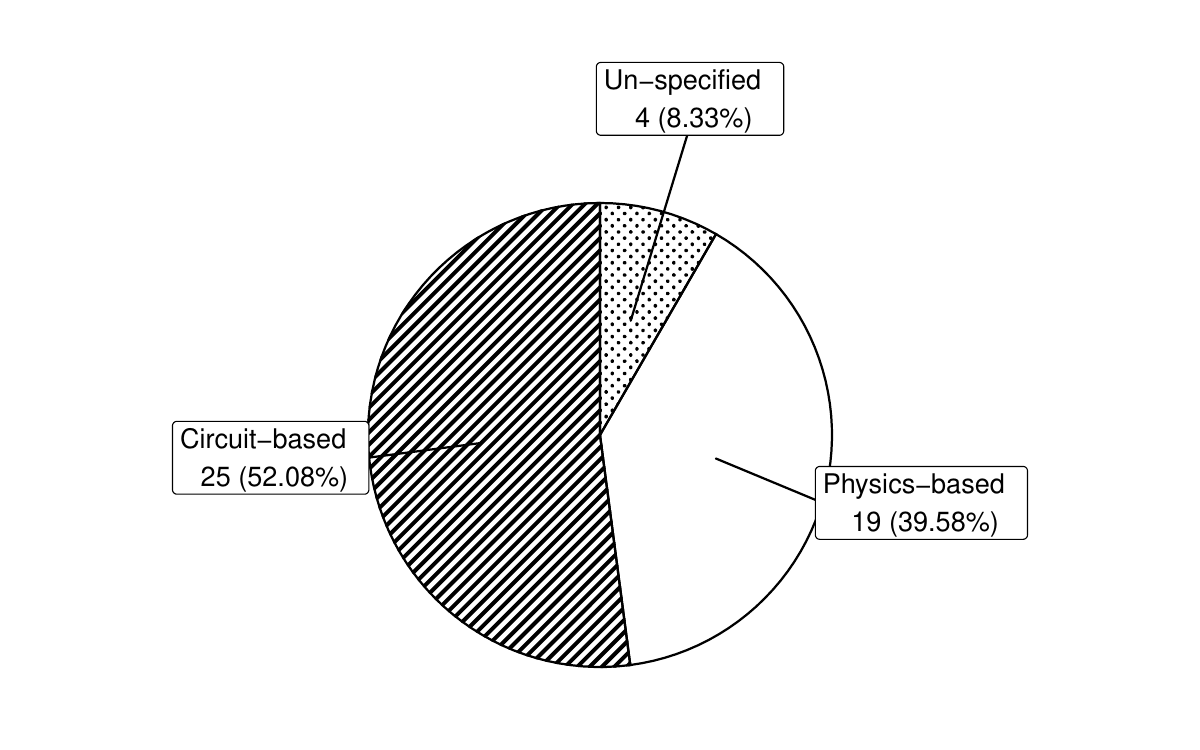}
	\caption{Number and proportion of quantum-inspired solutions by types of inspiration mechanisms}
	\label{fig:rq8_mechanims}
\end{figure}

As shown in Figure~\ref{fig:rq8_mechanims}, we classify the inspiration mechanisms of quantum-inspired optimization solutions into two categories: circuit-based and physics-based. Circuit-based mechanisms draw inspiration from quantum circuit models, including quantum gates, qubit rotations, and measurement operations. These approaches often simulate
behaviors found in gate-based quantum computing. For instance, primary studies~\cite{\papersixteenth, \paperfortysecond} propose to apply quantum bit encoding to ant colony positions, which leverage the computational principles of quantum bits and encoding. Physics-based mechanisms are inspired by fundamental quantum physical concepts such as wave functions, potential wells, or even effects like the Doppler effect. For instance, the Quantum PSO (QPSO) algorithm applied in primary study~\cite{\paperthirtyfirst} employs physics-based inspiration mechanism for controller placement problem, as QPSO draws inspiration from quantum mechanics concepts such as wave functions. From the results, we see that 52.08\% primary studies apply circuit-based inspiration mechanisms while 39.58\% primary studies are based on physics-based inspiration. 

\begin{table}
	\small
	\centering
	\caption{Summary of algorithms to extend or adopt, and mechanisms applied in quantum-inspired optimization}
	\label{tab:rq8_classical}
		\input{generated_files/RQ8_inspired_info.tex}
\end{table}

\begin{results}[Findings of RQ8]
	PSO and GA are the two foundational algorithms most commonly extended in quantum-inspired optimization, with proportions of 33.33\% and 16.67\%, respectively.
	Regarding PSO, it may lie in its inherent conceptual alignment with quantum mechanics, e.g., its particle-based representation.
	For GA, it may be due to its widespread application in SBSE.
	Among the 48 quantum-inspired solutions, 52.08\% are based on quantum circuit models, while 39.58\% are inspired by fundamental physical concepts.
\end{results}
\subsection{\rqCQOComputing}
\begin{figure}
	\centering
	\begin{subfigure}{0.45\textwidth}
		\includegraphics[width=\linewidth]{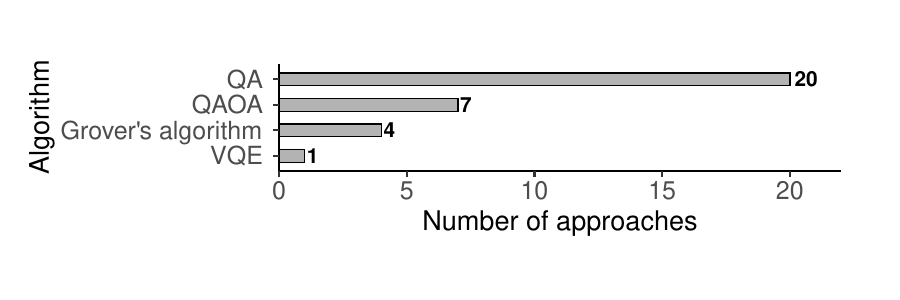}
		\caption{Algorithm}
		\label{subfig:rq9_algorithm}
	\end{subfigure}
	\begin{subfigure}{0.45\textwidth}
		\includegraphics[width=\linewidth]{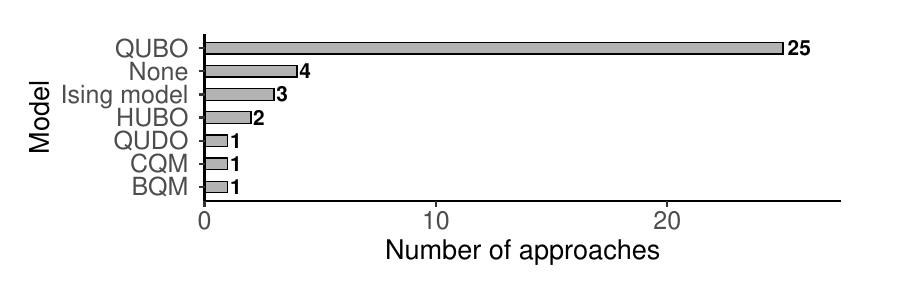}
		\caption{Mathematical Model}
		\label{subfig:rq9_mathmodel}
	\end{subfigure}
	\begin{subfigure}{0.45\textwidth}
		\includegraphics[width=\linewidth]{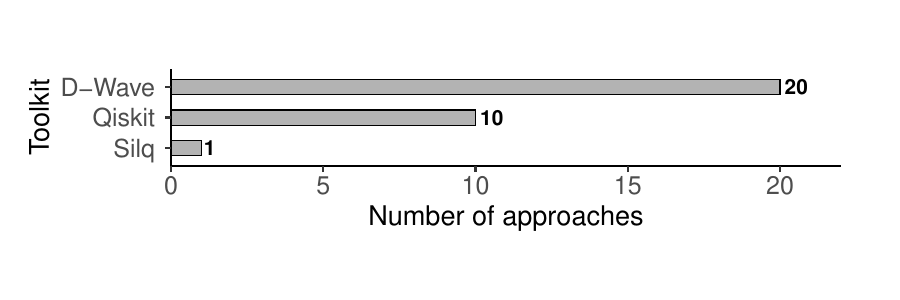}
		\caption{Programming Toolkit}
		\label{subfig:rq9_programming}
	\end{subfigure}
	\begin{subfigure}{0.45\textwidth}
		\includegraphics[width=\linewidth]{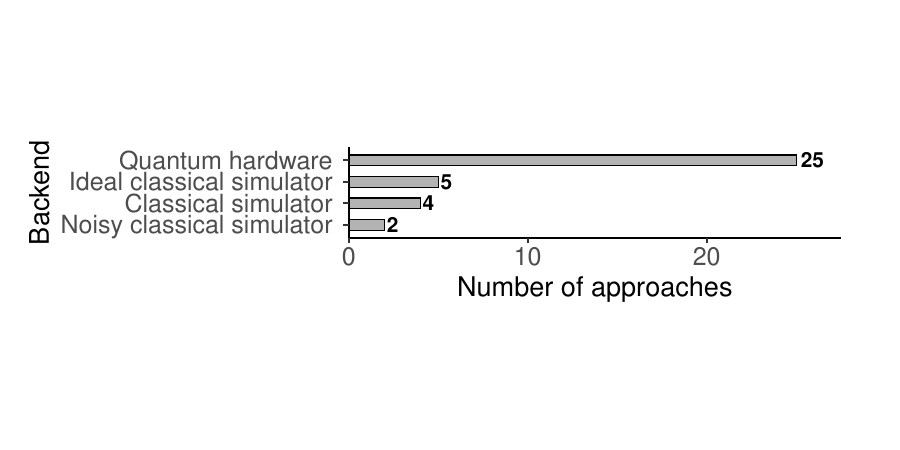}
		\caption{Backend}
		\label{subfig:rq9_backend}
	\end{subfigure}
	\caption{Frequency of quantum-related techniques employed in quantum approaches}
	\label{fig:rq9_quantum_solutions}
\end{figure}

This RQ focuses on quantum optimization, aiming to investigate which quantum algorithms have been applied, the mathematical models on which they are based, the programming platforms used for their execution, the computing backends employed, and the classical components integrated within hybrid approaches. We summarize the results in Table~\ref{tab:rq_quantum_solutions} and Figure~\ref{fig:rq9_quantum_solutions}.

First, we classify the quantum solutions proposed by the primary studies into two categories: hybrid and quantum. By ``hybrid'', we refer to approaches that combine quantum algorithms with classical components for problem decomposition, etc. By ``quantum'', we mean solutions that rely solely on pure quantum algorithms. 
From Table~\ref{tab:rq_quantum_solutions}, we can observe that among the 17 hybrid quantum optimization solutions and 15 purely quantum optimization solutions, four types of quantum algorithms have been applied: QA, QAOA, VQE, and Grover Search.QA, QAOA, and VQE directly target combinatorial or complex optimization problems, whereas Grover Search accelerates search-based subroutines in optimization workflows. 
Among these four, QA appears to be the most frequently applied quantum optimization algorithm. This may be due to the fact that QA is implemented on commercially available quantum annealers such as D-Wave devices, which are more mature and scalable for specific problems than gate-based quantum computers running QAOA and VQE. 
Furthermore, QA can be easily integrated with classical computing resources, such as decomposing large problems into QUBO subproblems and solve them with quantum annealers (e.g., ~\cite{\papersixtythird, \papersixtyfourth}). From Table~\ref{tab:rq_quantum_solutions}, we also see that all the primary studies that employ QA used D-Wave as the computing backend.
QAOA, as we discussed in Section~\ref{sec:background}, was designed to solve optimization problems using gate-based quantum computers, and is a hybrid quantum-classical algorithm, where a quantum circuit generates candidate solutions and a classical optimizer iteratively updates the circuit parameters to optimize the expected cost. As shown in Table~\ref{tab:rq_quantum_solutions}, all QAOA applications were implemented using the Qiskit programming platform,
 utilizing various backends including ideal simulators, noisy simulators, and real quantum hardware of IBM. 
VQE is also a hybrid quantum algorithm that uses a parameterized quantum circuit and classical optimization to solve optimization problems, which is only applied in primary study [73] on D-Wave's simulator applying simulated annealing. 
We consider Grover Search a fundamentally a quantum algorithm as its core research process runs on a quantum computer. As shown in Table~\ref{tab:rq_quantum_solutions}, four primary studies have applied Grover Search to solve different problems, mainly using Qiskit and Silq on various backends, including ideal simulators, noisy simulators, and quantum hardware.

As listed in Table~\ref{tab:rq_quantum_solutions}, the mathematical models used in quantum optimization and specified in the primary studies of our SLR include QUBO, BQM, HUBO, the Ising model, CQM, and QUDO. Among these, QUBO is the most widely adopted model and is frequently combined with quantum algorithms: QA, QAOA, and VQE. Other models like BQM, HUBO, CQM, and QUDO, which can be considered as diverse QUBO variants or extensions, are also paired with different quantum algorithms. This demonstrates diverse attempts to explore various problem formulations to improve solutions' cost-effectiveness. 

In quantum optimization solutions, classical components appear in three main forms: embedded at the platform level (e.g., classical optimizers involved in the hybrid solver of D-Wave), inherent to the algorithm itself (e.g., QAOA or QA serving as an optimizer of QBoost), or explicitly introduced (by users for instance) through hybrid designs, such as classical pre- or post-processing steps (e.g., warm-start and cold-start schemes~\cite{\papereightyseventh, \papereightyeighth}). These layers of classical integration reflect the practical need to complement current quantum capabilities with classical computation. For the case of only involving classical components at the platform level, we still classify them as pure-quantum categories. For the other two types, they belong to the hybrid category. We also observe cases that employ two ways of combining classical components. For instance, primary study~\cite{\papersixtyfourth} presents a hybrid approach that decomposes a problem to be solved into subproblems and then uses D-Wave to solve them. 

\begin{table}
	\small
	\centering
	\caption{Summary of quantum algorithms}
	\label{tab:rq_quantum_solutions}
	\resizebox{.99\textwidth}{!}{
		\input{generated_files/RQ9_quantum_info.tex}
			}
	\begin{spacing}{0.8}
		\raggedright \footnotesize 
		Herein, we offer the full names of the involved mathematical models: Binary Quadratic Model for BQM, Quadratic Unconstrained Binary Optimization for QUBO, Higher-order Unconstrained Binary Optimization for HUBO, and Quadratic Unconstrained Discrete Optimization for QUDO.
	\end{spacing}

\end{table}

\begin{results}[Findings of RQ9]
	Among the 32 solutions utilizing quantum algorithms, QA (with D-Wave) is the most commonly applied (20 out of 32 cases). QAOA is the second most used, appearing in 7 approaches. QUBO is the most widely adopted formulation, featured in 25 out of the 32 solutions, and is compatible with all quantum algorithms except for Grover’s algorithm.
\end{results}

\section{Identified Challenges}\label{sec:openChallenges}

In this section, we investigate the challenges of applying quantum optimization to SE identified in the literature (RQ10), and the remaining challenges yet to be addressed within the context of SDLC~(RQ11).

\subsection{\rqEClaimed} 
This RQ investigates what research challenges that have been explicitly discussed in the primary studies. After reviewing the related contexts from the collected papers, we summarize 12 labels that can represent the main issues of the discussed challenges. 

\begin{table}
	\small
	\centering
	\caption{Summary of identified challenges}
	\label{tab:rq14_identified_challenges}
	\resizebox{.99\textwidth}{!}{
		\input{generated_files/RQ14_identified_challenges.tex}
	}
\end{table}

We list the results in Table~\ref{tab:rq14_identified_challenges}. From the table, one can see that 29 primary studies mentioned the challenge of utilizing the involved approaches to address more complicated cases of problems investigated in their papers or other problems within SE (i.e., extensible approaches). Meanwhile, 27 primary studies concerned the challenge of exploring other approaches, including embedding them as a part of the proposed ones, to solve the same problems investigated in their papers (i.e., extensible problems). We also see 27 primary studies openly discussing how their approaches should be improved in terms of the structures, principles, and methodologies (i.e., solution improvement), which goes far away from merely tuning hyperparameters or adjusting configurations. Twenty primary studies discussed the challenge of enlarging the scalability of the adopted benchmarks (i.e., benchmark scalability), such as adopting programs with larger scales and databases with greater diversity.
Challenges in the NISQ era, such as the limited number of usable qubits and the negative impact of quantum noise on quantum devices, are also explicitly mentioned in 12 studies. Calling for real-world scenarios to evaluate proposed solutions has also been considered by 11 primary studies (i.e., real-world scenarios), as the empirical studies conducted in some of the primary studies only included artificial databases or benchmark testing. Seven primary studies underlined the need for parameter tuning or configuration optimization of the used approaches in their future works. Other concerns include the fault-tolerant technique used for the physical deployment of the quantum algorithms, the affordable overhead of implementing the optimization procedures, and the attempts on quantum hardware because of some empirical studies confined to classical simulators.


The top three most frequently identified challenges are extensible approaches, extensible problems, and solution improvement, indicating a strong need for scalable methods, broader problem applicability, and enhanced optimization quality.
This might suggest that quantum optimization techniques are still in their early stages of development and are drawing growing interest from both academia and industry.
The identified challenges of benchmark scalability and real-world scenarios are commonly recognized within the SE domain for conducting experiments to assess SE approaches.
From the perspective of quantum solutions, the most frequently identified challenges include the practical application of NISQ-era devices and effective parameter tuning.

\begin{results}[Findings of RQ10]
	The primary studies mostly consider that exploring different solutions for the same problem, discovering more difficult and real-world application scenarios of a given solution, and pursing large-scale, diverse problems and datasets are remaining open challenges. This is largely because the research field is still immature and requires further knowledge accumulation.
	Regarding the application of quantum techniques, the practical application of NISQ-era devices and effective parameter tuning are among the most frequently reported challenges. 
\end{results}

\subsection{\rqEOpen}

\subsubsection{SE Activities not yet covered according to SWEBOK}

First, we examine the SE activities that have been covered by the primary studies (see Figure~\ref{fig:rq4_se_activity}) and hence identify those not yet being targeted, by following the main chapters of SWEBOK. Results are displayed in Table~\ref{tab:coverage_for_SWEBOK}. 
Regarding ``Software Requirements'', we see that only two primary studies~\cite{\papertwentyeighth, \papertwentyninth} targeted the requirement selection problem (see Table~\ref{tab:rq5_se_problem}). Other optimization problems such as requirement prioritization~\cite{zhang2020uncertainty} have not yet been addressed by the literature. 
When looking at ``Software Architecture", as mentioned in SWEBOK, it is important to perform architecture tradeoff analysis to optimize software architectures based on their quality attributes. In the
past, search algorithms have been applied for architecture tradeoff analysis as summarized in~\cite{sobhy2021evaluation}. However, as shown in Figure~\ref{fig:rq4_se_activity}, there is no single primary study that targets software architecture optimization. 

Regarding ``Software Design', only the primary study~\cite{\paperninetieth} tackles the feature selection problem in recommendation systems. However, in this field, there exist other optimization problems, such as software design quality analysis which requires optimizing tradeoffs among modularity, portability, usability, and other factors~\cite{lu2017automated}.

As shown in Figure~\ref{fig:rq4_se_activity}, there is no primary study that targets any problem belonging to ``Software Construction''. However, various classical optimization algorithms have been applied for performance analysis and tuning~\cite{laaber2024evaluating},
code generation~\cite{gou2024rrgcode}, 
etc. 
Our SLR results show that there is great potential to apply quantum and quantum-inspired algorithms for addressing optimization problems in software design and construction. 

As shown in Figure~\ref{fig:rq4_se_activity}, ``Software Testing'' received the attention of 16 primary studies, mainly due to the fact that software testing is a highly studied field in classical software engineering~\cite{harman2012search}. 
As listed in SWEBOK, there are many testing activities that involve optimization, such as test generation and test optimization which have been investigated by the primary studies preliminarily. Many other challenges, such as scenario-based testing 
with search~\cite{lu2021search, zhu2023critical} and mutation testing with search~\cite{bottaci2001genetic, rahman2022review}, have not been addressed. We see a great opportunity to investigate how quantum and quantum-inspired algorithms can help improve the cost-effectiveness of testing approaches of classical software engineering. 

``Software Engineering Operations'' has been the subject that received extensive attention in the primary studies, as we discussed in Section~\ref{subsec:SEactivity}. We think it is because modern distributed systems, cloud environments, and large databases create optimization problems with exponentially large search spaces and classical algorithms struggle with scale to solve them. Hence researchers in these fields are more active in searching for new solutions such as quantum or quantum-inspired optimization, as compared to requirements engineering, design, etc. From Table~\ref{tab:rq5_se_problem}, we further observe that the problems belonging to SE operations are very diverse, including problems in database query optimization, service deployment, configuration optimization, resource optimization, node localization, communication optimization, etc., which are fundamental challenges in achieving the desired levels of performance, reliability, efficiency, and cost-effectiveness while the software is operating. We foresee that SE operations will continuously be an active field where quantum or quantum-inspired algorithms are applied in the future.

\begin{table}[!t]
	\small
	\centering
	\caption{Coverage of SE activities in SWEBOK}
	\label{tab:coverage_for_SWEBOK}
	\resizebox{.98\textwidth}{!}{
		\input{tables/coverage-swebok}
	}
	\begin{spacing}{0.8}
		\raggedright \footnotesize 
		Note that we only list the SWEBOK's chapters that correspond to concrete SE activities. 
	\end{spacing}
\end{table}

In terms of ``Software Maintenance'', according to our study, there is only one primary study~\cite{\paperfirst} about code cloning detection, which is formulated as a single-objective optimization problem and solved with QA of D-Wave. When referring to SWEBOK, we identify a great potential for applying quantum or quantum-inspired algorithms in this field. For instance, search-based software evolution~\cite{arcuri2009search} and search-based software maintenance~\cite{o2006search} have been well-studied by the SBSE community. We anticipate that as modern software systems grow in complexity (e.g., microservice architecture), optimizing software maintenance will become increasingly challenging. This necessitates exploring quantum solutions to address such challenges in the future.

There is only one primary study targeted ``Software Configuration Management''~\cite{\papersecond}. This primary study proposes to solve configuration selection and prioritization in the context of software product line engineering. Though this is the very first study, it demonstrates significant potential. Moreover, with the increasing complexity of highly configurable systems (e.g., cyber-physical systems), we believe there is great potential for applying quantum solutions in this field.
%
There is also one primary study for ``Software Engineering Management''~\cite{\paperfortyeighth}.
This primary study demonstrates the significant potential of applying quantum in this area, to address large-scale optimization problems in software project resource optimization, planning optimization, and even the Next Release Planning problem which has been well-studied by the SBSE community~\cite{xuan2012solving}.

There is no primary study targeted on ``Software Engineering Process'' topics, which are mostly about SDLC and process assessment and improvement. In the history of SBSE, this topic is also relatively less studied~\cite{harman2012search}.

As discussed in Section~\ref{subsec:SEactivity}, ``Software Quality'' attracted 10 primary studies, 7 of which are about optimization in software failure prediction using quantum and quantum-inspired solutions. We see significant opportunities to apply these approaches to other challenging software quality tasks, such as CI/CD pipeline optimization and technical debt management. Software quality directly determines system resilience, cost efficiency, and user trust; hence, it is important to continuously look for novel solutions to improve software quality.

Among the \finalselected primary studies, five of them are about ``Software Security'' (Figure~\ref{fig:rq4_se_activity}), and all focus on security attack identification (Table~\ref{tab:rq5_se_problem}). Other critical optimization opportunities in vulnerability prevention and security policy enforcement where quantum solutions could offer promising alternatives yet to be discovered in the future. 

%
Other chapters in SWEBOK, e.g., ``Software Engineering Professional Practice'' and ``Software Engineering Models and Methods'' are meta-level concerns decoupled from concrete SDLC activities; hence, no primary study was mapped to these domains, as current quantum and quantum-inspired optimization prioritizes phase-specific technical optimizations.

\subsubsection{Mapping of solutions to types of problems}
\begin{figure}
	\centering
	\includegraphics[width=.9\textwidth]{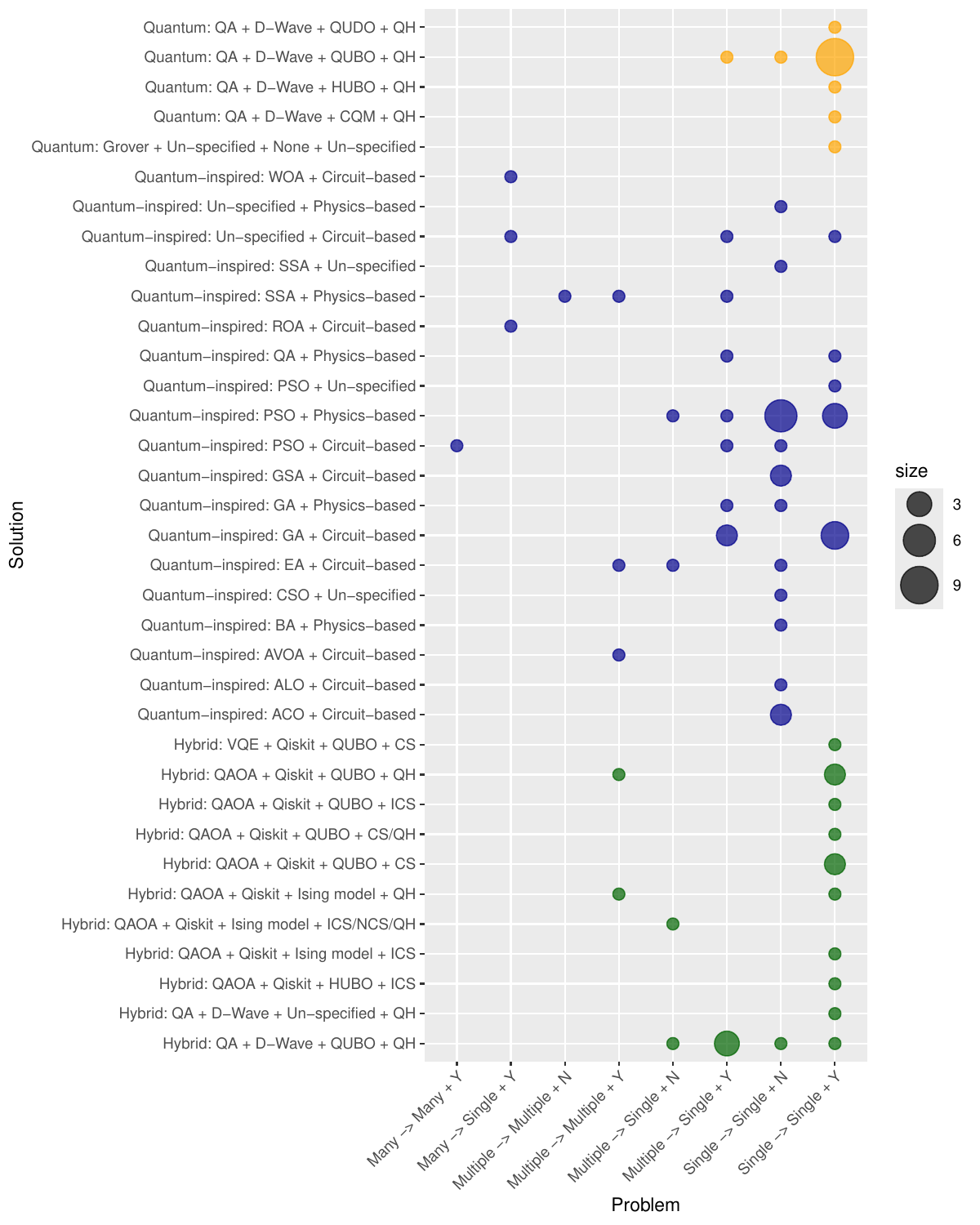}
	\caption{Types of problems (x-axis) addressed by each kind of quantum solutions (y-axis)}
	\label{fig:rq11_solution_summary}
\end{figure}

In this section, we examine the mapping between specific quantum solutions and the characteristics of the problems and its formulations. To this end, we plot Figure~\ref{fig:rq11_solution_summary}, were each quantum solution is associated to categories of optimization problems, with the bubble size indicating the number of primary studies that apply this quantum solution for solving this category of the problems. 
The categories are primarily defined based on the objectives of the SE problems and their optimization reformulations. 
For example,``Multiple $\rightarrow$ Single + Y'' indicates that a constrained multi-objective SE problem has been reformulated as a constrained single-objective problem, which is then addressed using quantum optimization techniques (y-axis).
These techniques on the y-axis are denoted as \textit{<Type>: <Base Algorithm> + <Mechanism>} for quantum-inspired approaches, and as \textit{<Type>: <Quantum Algorithm> + <Programming Platform> + <Mathematical Model> + <Computing Backend>} for quantum and hybrid approaches.

In terms of problem categories, many- and multi-objective problems are less frequently addressed in quantum and quantum-inspired optimization approaches. 
However, in the field of SBSE, many SE problems have been reformulated as many- and multi-objective search problems to be effectively tackled by search techniques~\cite{harman2012search, ramirez2018systematic}.
This reveals significant untapped potential for quantum optimization to contribute in this area.

Regarding quantum-inspired approaches, as previously discussed, PSO and GA are the most commonly extended algorithms for tackling SE problems, which are often reformulated into single-objective formulations for solving. 
This presents an opportunity to investigate whether popular many- and multi-objective algorithms (such as NSGA-II, NSGA-III, SPEA2) can be feasibly extended using quantum mechanisms.
Regarding quantum solutions, all employ QA and are currently developed using D-Wave systems and quantum hardware. 
Exploring additional algorithms and hardware platforms offers significant potential for future research.
Regarding hybrid solutions, QAOA shows potential to address multi-objective problems.
This implies that it may also be extended to tackle many-objective problems and be applied to a broader range of SE challenges.
Regarding quantum and hybrid solutions, QUBO is the dominant model used to reformulate SE problems for solving with quantum techniques. 
Models such as HUBO, QUDO and CQM are emerging in the field, with more studies needed to explore their potential applications.


For solving single-objective SE problems with constraints, D-Wave (with QA and QUBO) is the most used quantum solution (i.e., ``Quantum: QA + D-Wave + QUDO + QH''), followed by quantum-circuit-inspired GA and quantum-physics-inspired PSO. For solving single-objective with no constraints (the simplest optimization problem), quantum-physics-inspired PSO outperformed all the others, and almost all solutions are quantum-inspired solutions. For the multi-objective SE problems that were eventually converted to single-objective problems with constraints, D-Wave is still the first option but combined with classical components (e.g., decomposition of a large problem into sub-problems), which are hence hybrid (i.e., ``Hybrid: QA + D-Wave + QUBO + QH''). For many and multi-objective SE problems were more often solved with quantum-inspired solutions. Due to the lack of data points, we cannot perform analyses for each many and multi-objectives problems, unfortunately. More details are presented in  Table~\ref{tab:ComparingOptimizationMethods} for reference.

To further know how a specific SE problem being solved by different solutions, we present Table~\ref{tab:rq_solution_summary_bySEproblem}. 
As shown in the table, all scheduling problems were solved by quantum-inspired solutions, among which PSO was mostly selected as the base algorithm. 
For solving test suite minimization problems, most of the primary studies opted for quantum-inspired solutions;  three different types quantum solutions have been experimented, e.g., QA, QAOA and Grover Search; and both single-objective and multi-objective problems were converted into single-objective QUBO formulations. 


\begin{results}[Findings of RQ11]
	By following SE activities defined in SWEBOK, six out of 15 SE activities have been covered by the primary studies, i.e., Software Architecture, Software Construction, SE Process, SE Models and Methods, SE Professional Practice, and SE Economics.
	Among the covered SE activities, there are a range of tasks not yet been targeted by the literature. 
	Many- and multi-objective problems are less addressed in quantum and quantum-inspired optimization approaches, indicating significant untapped potential for quantum optimization to contribute in this area.
\end{results}

%% file: generated_files/RQ2_venue.tex
\begin{tabular} { l p{0.3\textwidth} r r r} \\ 
\toprule 
 & Venue &  Number of Papers & Venue Type & SE/Non-SE\\ 
\midrule  
1 & \emph{Other Journal} & 29 & Journal & Non-SE, SE\\ 
2 & \emph{Other Conference} & 16 & Conference & Non-SE, SE\\ 
3 & \emph{IEEE International Conference on Quantum Computing and Engineering (QCE)} & 5 & Conference & Non-SE\\ 
4 & \emph{Other Workshop} & 4 & Workshop & Non-SE, SE\\ 
5 & \emph{Applied Soft Computing} & 3 & Journal & Non-SE\\ 
6 & \emph{Arxiv} & 3 & Open-Access Archive & Non-SE\\ 
7 & \emph{Cluster Computing} & 3 & Journal & Non-SE\\ 
8 & \emph{IEEE Access} & 3 & Journal & Non-SE\\ 
9 & \emph{Sensors} & 3 & Journal & Non-SE\\ 
10 & \emph{ACM Transactions on Software Engineering and Methodology (TOSEM)} & 2 & Journal & SE\\ 
11 & \emph{IEEE Transactions on Software Engineering (TSE)} & 2 & Journal & SE\\ 
12 & \emph{IET Software} & 2 & Journal & SE\\ 
13 & \emph{International Workshop on Quantum Programming for Software Engineering (QP4SE)} & 2 & Workshop & SE\\ 
14 & \emph{Soft Computing} & 2 & Journal & Non-SE\\ 
15 & \emph{Workshop on Quantum Computing and Quantum-Inspired Technology for Data-Intensive Systems and Applications (Q-Data)} & 2 & Workshop & Non-SE\\ 
\bottomrule 
\end{tabular} 

%% file: generated_files/RQ5_problem.tex
\begin{tabular} {l l l l l} \\ 
\toprule 
SE Activity & No.& SE Problem & Papers\\ 
\midrule  
software engineering operations & 1 & \textcolor{green}{$\blacktriangleright$} \emph{scheduling problem} & 8: \cite{ \paperthird,\paperseventh,\paperfourteenth,\papertwentyfourth,\paperfortyfirst,\paperseventysecond,\papereightysecond,\paperninetyfirst }\\ 
 & 2 & \emph{join order problem} & 6: \cite{ \paperfifth,\paperfiftieth,\paperfiftysecond,\paperfiftyfourth,\papersixtyfirst,\paperninetythird }\\ 
 & 3 & \emph{allocation problem} & 4: \cite{ \paperseventieth,\papereightysixth,\papereightyseventh,\papereightyeighth }\\ 
 & 4 & \emph{controller placement} & 3: \cite{ \paperthirtyfirst,\paperfortyfourth,\papereightythird }\\ 
 & 5 & \emph{routing optimization} & 3: \cite{ \papereightyfourth,\papereightyninth,\paperninetysecond }\\ 
 & 6 & \textcolor{blue}{$\blacktriangleright$} \emph{security attack identification} & 3: \cite{ \papertwentysecond,\paperfiftyfifth,\papersixtyninth }\\ 
 & 7 & \emph{multiple query optimization} & 2: \cite{ \paperfiftythird,\paperfiftyeighth }\\ 
 & 8 & \textcolor{orange}{$\blacktriangleright$} \emph{security management} & 2: \cite{ \papertwentyfifth,\papersixtyeighth }\\ 
 & 9 & \emph{service deployment} & 2: \cite{ \paperthirtysixth,\paperthirtyninth }\\ 
 & 10 & \emph{clustering protocol optimization} & 1: \cite{ \papereightyfifth }\\ 
 & 11 & \emph{communication optimization} & 1: \cite{ \paperthirtyfifth }\\ 
 & 12 & \emph{container management} & 1: \cite{ \paperfortythird }\\ 
 & 13 & \emph{database index selection} & 1: \cite{ \papersixtieth }\\ 
 & 14 & \emph{network delay optimization} & 1: \cite{ \papereightieth }\\ 
 & 15 & \emph{node localization} & 1: \cite{ \papertwentythird }\\ 
 & 16 & \emph{queuing delay optimization} & 1: \cite{ \paperseventyninth }\\ 
 & 17 & \emph{resource optimization} & 1: \cite{ \paperfortyfifth }\\ 
 & 18 & \emph{service configuration} & 1: \cite{ \paperfortieth }\\ 
 & 19 & \textcolor{red}{$\blacktriangleright$} \emph{software failure prediction} & 1: \cite{ \papersixtyseventh }\\ 
\midrule  
software testing & 1 & \emph{test suite minimization} & 9: \cite{ \papertenth,\paperthirteenth,\papersixteenth,\paperthirtythird,\paperthirtyseventh,\paperthirtyeighth,\papersixtyfourth,\paperseventyfourth,\paperseventyseventh }\\ 
 & 2 & \emph{covering array generation} & 5: \cite{ \papertwentieth,\paperthirtyfourth,\papersixtysixth,\paperseventyfifth,\paperseventysixth }\\ 
 & 3 & \emph{test case prioritization} & 3: \cite{ \papertenth,\paperthirtythird,\paperthirtyeighth }\\ 
 & 4 & \emph{test case selection} & 3: \cite{ \paperthirteenth,\paperthirtyeighth,\papersixtythird }\\ 
 & 5 & \emph{dynamic testing} & 1: \cite{ \papersixtyfifth }\\ 
\midrule  
software quality & 1 & \textcolor{red}{$\blacktriangleright$} \emph{software failure prediction} & 7: \cite{ \papereleventh,\papertwelfth,\papertwentysixth,\paperthirtieth,\papersixtyseventh,\paperseventyfirst,\paperseventythird }\\ 
 & 2 & \emph{DTC problem} & 1: \cite{ \paperfourth }\\ 
 & 3 & \emph{software reliability growth model} & 1: \cite{ \paperthirtysecond }\\ 
 & 4 & \emph{WCET evaluation} & 1: \cite{ \papertwentyseventh }\\ 
\midrule  
software security & 1 & \textcolor{blue}{$\blacktriangleright$} \emph{security attack identification} & 6: \cite{ \papersixth,\papertwentysecond,\paperfortysecond,\paperfiftyfifth,\paperfiftysixth,\papersixtyninth }\\ 
 & 2 & \textcolor{orange}{$\blacktriangleright$} \emph{security management} & 2: \cite{ \papertwentyfifth,\papersixtyeighth }\\ 
\midrule  
software requirements & 1 & \emph{software requirement selection} & 2: \cite{ \papertwentyeighth,\papertwentyninth }\\ 
\midrule  
software configuration management & 1 & \emph{configuration prioritization} & 1: \cite{ \papersecond }\\ 
 & 2 & \emph{configuration selection} & 1: \cite{ \papersecond }\\ 
\midrule  
software design & 1 & \emph{feature selection} & 1: \cite{ \paperninetieth }\\ 
\midrule  
software engineering management & 1 & \textcolor{green}{$\blacktriangleright$} \emph{scheduling problem} & 1: \cite{ \paperfortyeighth }\\ 
\midrule  
software maintainance & 1 & \emph{code clone detection} & 1: \cite{ \paperfirst }\\ 
\bottomrule 
\end{tabular} 

%% file: generated_files/RQ6_problem_reformulation.tex
\begin{tabular} {p{0.1\textwidth} p{0.15\textwidth} p{0.12\textwidth} p{0.4\textwidth} p{0.18\textwidth}} \\ 
\toprule 
\# of Obj. & \# of Ref. Obj. & Constraints & SE Problem & Papers \\ 
\midrule  
Many & Many&Y&Scheduling problem&1:\cite{ \paperseventysecond }\\ 
 & Single&Y&Resource optimization, Clustering protocol optimization, Routing optimization&3:\cite{ \paperfortyfifth,\papereightyfifth,\papereightyninth }\\ 
\midrule  
Multiple & Multiple&N&Controller placement&1:\cite{ \papereightythird }\\ 
 & Multiple&Y&Software requirement selection, Software requirement selection, Controller placement, Routing optimization&4:\cite{ \papertwentyeighth,\papertwentyninth,\paperfortyfourth,\papereightyfourth }\\ 
 & Single&N&Test case selection, Test suite minimization, Test suite minimization, Allocation problem, Routing optimization&4:\cite{ \paperthirteenth,\papersixtyfourth,\paperseventieth,\paperninetysecond }\\ 
 & Single&Y&Scheduling problem, Service configuration, Scheduling problem, Scheduling problem, Test case selection, Security management, Queuing delay optimization, Scheduling problem, Allocation problem, Allocation problem, Allocation problem&11:\cite{ \paperthird,\paperfortieth,\paperfortyfirst,\paperfortyeighth,\papersixtythird,\papersixtyeighth,\paperseventyninth,\papereightysecond,\papereightysixth,\papereightyseventh,\papereightyeighth }\\ 
\midrule  
Single & Single&N&Test suite minimization, Test case prioritization, Software failure prediction, Software failure prediction, Test suite minimization, Covering array generation, Security attack identification, Node localization, Software failure prediction, Software reliability growth model, Test case prioritization, Test suite minimization, Test case prioritization, Test case selection, Test suite minimization, Security attack identification, Security attack identification, Covering array generation, Software failure prediction, Software failure prediction, Software failure prediction, Test suite minimization, Covering array generation&19:\cite{ \papertenth,\papereleventh,\papertwelfth,\papersixteenth,\papertwentieth,\papertwentysecond,\papertwentythird,\paperthirtieth,\paperthirtysecond,\paperthirtythird,\paperthirtyeighth,\paperfortysecond,\paperfiftyfifth,\papersixtysixth,\papersixtyseventh,\paperseventyfirst,\paperseventythird,\paperseventyfourth,\paperseventysixth }\\ 
 & Single&Un-specified&Security attack identification, Container management, Security attack identification&3:\cite{ \papersixth,\paperfortythird,\papersixtyninth }\\ 
 & Single&Y&Code clone detection, Configuration selection, Configuration prioritization, Join order problem, Security management, Software failure prediction, WCET evaluation, Controller placement, Covering array generation, Service deployment, Test suite minimization, Service deployment, Join order problem, Join order problem, Multiple query optimization, Join order problem, Security attack identification, Multiple query optimization, Database index selection, Join order problem, Covering array generation, Test suite minimization, Network delay optimization, Routing optimization, Feature selection, Scheduling problem, Join order problem&26:\cite{ \paperfirst,\papersecond,\paperfifth,\papertwentyfifth,\papertwentysixth,\papertwentyseventh,\paperthirtyfirst,\paperthirtyfourth,\paperthirtysixth,\paperthirtyseventh,\paperthirtyninth,\paperfiftieth,\paperfiftysecond,\paperfiftythird,\paperfiftyfourth,\paperfiftysixth,\paperfiftyeighth,\papersixtieth,\papersixtyfirst,\paperseventyfifth,\paperseventyseventh,\papereightieth,\papereightyfourth,\paperninetieth,\paperninetyfirst,\paperninetythird }\\ 
\bottomrule 
\end{tabular} 

%% file: generated_files/RQ7_quantum_type.tex
\begin{tabular} { l r p{0.5\textwidth} } \\ 
\toprule 
 Type & \#  & Papers \\ 
\midrule  
Classical & 48 & \cite{ \paperthird,\paperfifth,\papertenth,\papertwelfth,\paperfourteenth,\papersixteenth,\papertwentieth,\papertwentysecond,\papertwentythird,\papertwentyfourth,\papertwentysixth,\papertwentyeighth,\papertwentyninth,\paperthirtieth,\paperthirtyfirst,\paperthirtysecond,\paperthirtythird,\paperthirtyfourth,\paperthirtyfifth,\paperthirtysixth,\paperthirtyeighth,\paperthirtyninth,\paperfortieth,\paperfortyfirst,\paperfortysecond,\paperfortythird,\paperfortyfourth,\paperfortyfifth,\paperfortyeighth,\papersixtysixth,\papersixtyseventh,\paperseventieth,\paperseventyfirst,\paperseventysecond,\paperseventythird,\paperseventyfourth,\paperseventyfifth,\paperseventysixth,\paperseventyseventh,\paperseventyninth,\papereightieth,\papereightysecond,\papereightythird,\papereightyfifth,\papereightysixth,\papereightyninth,\paperninetyfirst,\paperninetysecond }\\ 
Hybrid & 16 & \cite{ \papersecond,\papersixth,\paperthirteenth,\papertwentyseventh,\paperfiftyfourth,\paperfiftyfifth,\paperfiftysixth,\paperfiftyeighth,\papersixtieth,\papersixtythird,\papersixtyfourth,\papersixtyninth,\papereightyfourth,\papereightyseventh,\papereightyeighth,\paperninetythird }\\ 
Quantum & 15 & \cite{ \paperfirst,\paperfourth,\paperseventh,\papereleventh,\papertwentyfifth,\papertwentyseventh,\paperthirtyseventh,\paperfiftieth,\paperfiftysecond,\paperfiftythird,\paperfiftyfourth,\papersixtyfifth,\papersixtyeighth,\paperninetieth,\paperninetythird }\\ 
\bottomrule 
\end{tabular} 

%% file: generated_files/RQ8_inspired_info.tex
\begin{tabular} {  l l l} \\ 
\toprule 
 Base Algorithm (\#, \%) & Mechanism & Papers \\ 
\midrule  
PSO (16, 33.33\%)&Circuit-based&3:\cite{ \papertwentieth,\paperseventysecond,\papereightysecond }\\ 
&Physics-based&11:\cite{ \paperthird,\papertenth,\paperthirtieth,\paperthirtyfirst,\paperthirtysecond,\paperthirtyfourth,\paperthirtyeighth,\paperseventieth,\paperseventyfirst,\paperseventyfifth,\paperseventysixth }\\ 
&\emph{Un-specified}&2:\cite{ \paperfourteenth,\papertwentysixth }\\ 
GA (8, 16.67\%)&Circuit-based&6:\cite{ \paperthirtysixth,\paperthirtyninth,\paperfortieth,\paperseventyseventh,\paperseventyninth,\papereightieth }\\ 
&Physics-based&2:\cite{ \paperthird,\papertenth }\\ 
SSA (4, 8.33\%)&Physics-based&3:\cite{ \paperfortyfirst,\paperfortyfourth,\papereightythird }\\ 
&\emph{Un-specified}&1:\cite{ \papertwentythird }\\ 
EA (3, 6.25\%)&Circuit-based&3:\cite{ \papertwentyeighth,\papersixtysixth,\paperninetysecond }\\ 
GSA (3, 6.25\%)&Circuit-based&3:\cite{ \paperthirtyfifth,\papersixtyseventh,\paperseventythird }\\ 
ACO (2, 4.17\%)&Circuit-based&2:\cite{ \papersixteenth,\paperfortysecond }\\ 
QA (2, 4.17\%)&Physics-based&2:\cite{ \paperfifth,\paperfortyeighth }\\ 
WOA (2, 4.17\%)&Circuit-based&1:\cite{ \papereightyfifth }\\ 
&Physics-based&1:\cite{ \papertwentyfourth }\\ 
ALO (1, 2.08\%)&Circuit-based&1:\cite{ \papertwentysecond }\\ 
AVOA (1, 2.08\%)&Circuit-based&1:\cite{ \papertwentyninth }\\ 
BA (1, 2.08\%)&Physics-based&1:\cite{ \paperthirtythird }\\ 
CSO (1, 2.08\%)&\emph{Un-specified}&1:\cite{ \papertwelfth }\\ 
RNN (1, 2.08\%)&Circuit-based&1:\cite{ \paperfortythird }\\ 
ROA (1, 2.08\%)&Circuit-based&1:\cite{ \papereightyninth }\\ 
\emph{Un-specified} (4, 8.33\%)&Circuit-based&3:\cite{ \paperfortyfifth,\papereightysixth,\paperninetyfirst }\\ 
&Physics-based&1:\cite{ \paperseventyfourth }\\ 
\bottomrule 
\end{tabular} 

%% file: generated_files/RQ9_quantum_info.tex
\begin{tabular} { l p{0.1\textwidth} p{0.12\textwidth} p{0.12\textwidth} p{0.22\textwidth} p{0.2\textwidth} p{0.15\textwidth}} \\ 
\toprule 
Quantum Type & Quantum Algorithm & Mathematical Model & Programming Platform & Computing Backend & Classical Component & Papers \\ 
\midrule  
Hybrid & QA&BQM&D-Wave&Quantum hardware&Platform, Algorithm-inherent&1:\cite{ \papersixth }\\ 
 & QA&QUBO&D-Wave&Quantum hardware&Algorithm-inherent&1:\cite{ \paperfiftyfifth }\\ 
 & QA&QUBO&D-Wave&Quantum hardware&Platform, Algorithm-inherent&1:\cite{ \papersixtyninth }\\ 
 & QA&QUBO&D-Wave&Quantum hardware&User-side&1:\cite{ \papersixtieth }\\ 
 & QA&QUBO&D-Wave&Quantum hardware&User-side, Platform&4:\cite{ \papersixtythird,\papersixtyfourth,\papereightyseventh,\papereightyeighth }\\ 
 & QA&Un-specified&D-Wave&Quantum hardware&Platform, Algorithm-inherent&1:\cite{ \paperfiftysixth }\\ 
 & QAOA&HUBO, QUBO, Ising model&Qiskit&Ideal classical simulator&Algorithm-inherent&1:\cite{ \papersecond }\\ 
 & QAOA&Ising model&Qiskit&Ideal classical simulator, Noisy classical simulator, Quantum hardware&User-side, Algorithm-inherent&1:\cite{ \paperthirteenth }\\ 
 & QAOA&QUBO&Qiskit&Classical simulator&Algorithm-inherent&2:\cite{ \papertwentyseventh,\paperfiftyfourth }\\ 
 & QAOA&QUBO&Qiskit&Classical simulator, Quantum hardware&Algorithm-inherent&1:\cite{ \paperfiftyeighth }\\ 
 & QAOA&QUBO&Qiskit&Quantum hardware&Algorithm-inherent&1:\cite{ \paperninetythird }\\ 
 & QAOA&QUBO, Ising model&Qiskit&Quantum hardware&Algorithm-inherent&1:\cite{ \papereightyfourth }\\ 
 & VQE&QUBO&Qiskit&Classical simulator&Algorithm-inherent&1:\cite{ \paperfiftyfourth }\\ 
\midrule  
Quantum & Grover&None&Qiskit&Ideal classical simulator&None&1:\cite{ \paperfourth }\\ 
 & Grover&None&Qiskit&Ideal classical simulator, Noisy classical simulator, Quantum hardware&Un-specified&1:\cite{ \papersixtyfifth }\\ 
 & Grover&None&Silq&Ideal classical simulator&None&1:\cite{ \paperseventh }\\ 
 & Grover&None&Un-specified&Un-specified&None&1:\cite{ \paperthirtyseventh }\\ 
 & QA&CQM, QUBO&D-Wave&Quantum hardware&Platform&1:\cite{ \paperfiftieth }\\ 
 & QA&HUBO, QUBO&D-Wave&Quantum hardware&Platform&1:\cite{ \paperfiftysecond }\\ 
 & QA&QUBO&D-Wave&Quantum hardware&Platform&4:\cite{ \papereleventh,\papertwentyfifth,\papersixtyeighth,\paperninetieth }\\ 
 & QA&QUBO&D-Wave&Quantum hardware&Un-specified&4:\cite{ \papertwentyseventh,\paperfiftythird,\paperfiftyfourth,\paperninetythird }\\ 
 & QA&QUBO, QUDO&D-Wave&Quantum hardware&Platform&1:\cite{ \paperfirst }\\ 
\bottomrule 
\end{tabular} 

%% file: generated_files/RQ14_identified_challenges.tex
\begin{tabular} { l l l} \\ 
\toprule 
 Identified Challenges & \# & Papers\\ 
\midrule  
\emph{Extensible approaches} & 29 & \cite{ \papersixth,\papertenth,\papereleventh,\papersixteenth,\papertwentieth,\papertwentysecond,\papertwentythird,\papertwentyfifth,\papertwentyeighth,\papertwentyninth,\paperthirtieth,\paperthirtyfirst,\paperthirtythird,\paperthirtyfifth,\paperthirtyeighth,\paperfiftysecond,\paperfiftythird,\paperfiftyfifth,\paperfiftysixth,\papersixtyfirst,\papersixtythird,\papersixtysixth,\papersixtyseventh,\papersixtyeighth,\papersixtyninth,\paperseventysecond,\papereightysecond,\papereightythird,\paperninetythird }\\ 
\emph{Extensible problems} & 27 & \cite{ \paperfirst,\paperseventh,\paperthirteenth,\paperfourteenth,\papersixteenth,\papertwentysecond,\papertwentyeighth,\paperthirtieth,\paperthirtysixth,\paperthirtyseventh,\paperthirtyninth,\paperfortyfirst,\paperfortythird,\paperfortyfourth,\paperfortyfifth,\paperfiftieth,\paperfiftythird,\paperfiftyfifth,\papersixtyfourth,\papersixtyeighth,\papersixtyninth,\paperseventieth,\papereightyfifth,\papereightysixth,\papereightyeighth,\paperninetyfirst,\paperninetysecond }\\ 
\emph{Solution improvement} & 27 & \cite{ \papersecond,\paperfourth,\paperseventh,\papertenth,\paperfourteenth,\papertwentieth,\papertwentyninth,\paperthirtythird,\paperthirtyfourth,\paperthirtyfifth,\paperthirtyseventh,\paperthirtyeighth,\paperfortieth,\paperfortyfifth,\paperfiftieth,\paperfiftysecond,\paperfiftyfifth,\papersixtyfirst,\papersixtythird,\papersixtyfourth,\papersixtyfifth,\papersixtysixth,\papersixtyseventh,\paperseventyfourth,\paperseventyfifth,\paperseventysixth,\paperninetysecond }\\ 
\emph{Benchmark scalability} & 20 & \cite{ \paperfirst,\paperfifth,\papereleventh,\paperthirteenth,\papertwentieth,\papertwentyfifth,\papertwentyninth,\paperthirtythird,\paperthirtyfifth,\paperthirtyeighth,\paperfiftythird,\paperfiftyfifth,\paperfiftysixth,\papersixtyfirst,\papersixtyfourth,\papersixtyfifth,\papersixtysixth,\paperseventyfourth,\papereightyfifth,\paperninetythird }\\ 
\emph{NISQ era} & 12 & \cite{ \paperfirst,\paperfourth,\paperthirteenth,\papertwentyseventh,\paperfiftythird,\paperfiftyfourth,\paperfiftyeighth,\papersixtyfirst,\papersixtyfourth,\papersixtyfifth,\papereightyfourth,\paperninetythird }\\ 
\emph{Real-world scenarios} & 11 & \cite{ \papersixth,\papertenth,\papertwentyeighth,\papertwentyninth,\paperthirtythird,\paperthirtyeighth,\paperfortyfourth,\papersixtyseventh,\papersixtyeighth,\paperseventyfourth,\papereightythird }\\ 
\emph{Parameter tuning} & 7 & \cite{ \papersecond,\papertwentysixth,\papertwentyseventh,\paperfortysecond,\paperfiftyfourth,\papersixtyfourth,\paperseventyfourth }\\ 
\emph{Automation} & 1 & \cite{ \papersixth }\\ 
\emph{Fault tolerance} & 1 & \cite{ \paperfourth }\\ 
\emph{Overhead} & 1 & \cite{ \paperfifth }\\ 
\emph{Quantum hardware backend} & 1 & \cite{ \papertwentyseventh }\\ 
\bottomrule 
\end{tabular} 

%% file: tables/coverage-swebok.tex
\begin{tabular}{ l l l | l l l } 
    \toprule 
    Chapter   & Title     & Covered?
    &  Chapter   & Title     & Covered? \\ 
    \midrule
    1       & Software Requirements            &    \markyes    &
    9       & Software Engineering Management  &    \markyes    \\
    2       & Software Architecture            &    \markno     &
    10      & Software Engineering Process     &    \markno     \\
    3       & Software Design                  &    \markyes    &
    11      & Software Engineering Models and Methods &    \markno     \\
    4       & Software Construction            &    \markno     &
    12      & Software Quality                 &    \markyes    \\
    5       & Software Testing                 &    \markyes    &
    13      & Software Security                &    \markyes    \\
    6       & Software Engineering Operations  &    \markyes    &
    14      & Software Engineering Professional Practice &    \markno     \\
    7       & Software Maintenance             &    \markyes    &
    15      & Software Engineering Economics   &    \markno     \\
    8       & Software Configuration Management&    \markyes    & & & \\
    \bottomrule
\end{tabular} 

%% file: tables/related-qse.tex
\begin{tabular}{l p{0.1\textwidth} l p{0.1\textwidth} p{0.35\textwidth} p{0.45\textwidth}} 
    \toprule 
    Works & Publication Year & Study Duration & \# of Primiary Studies & Phase and Area  & Research Method and Scope \\ 
    \midrule 
    ~\cite{rw-mandal-qse_qml}
    & 2025
    & 2004$\sim$2023 
    & 56 
    & Requirements, modeling, formal verification, software testing and debugging, refactoring, reverse engineering, empirical studies, and benchmarking
    & Via SLR, examined existing QSE techniques, revealed the potential of applying quantum machine learning and quantum optimization in classical software engineering, and proposed a research agenda for QSE. 
    \\
    \midrule 
    ~\cite{rw-Murillo-qse}  
    & 2025 
    & 2002$\sim$2023
    & 101
    & Software testing and debugging, service-oriented computing, model-driven engineering, programming paradigms, software architectures, software development processes, artificial intelligence  
    & Via SLR, analyzed active researchers in QSE and identified key QSE research areas, open challenges and future directions.\\
    \midrule 
    ~\cite{rw-Sep-requirements}
    & 2024
    & 2017$\sim$2024 
    & 105
    & Requirements engineering               
    & Via SLR, identified challenges of requirements engineering for quantum computing.\\
    \midrule  
    ~\cite{rw-Paltenghi-testing_analysis} 
    & 2024
    & 2007$\sim$2023 
    & 102 
    & Software testing and debugging, static analysis, and program optimization 
    & Via SLR, examined the literature spanning quantum computing, software engineering, programming languages, and formal methods.\\
    \midrule 
    ~\cite{rw-Garcia-testing}
    & 2023
    & 2004$\sim$2021 
    & 30 
    & Software testing and debugging
    & Via a systematic mapping study, identified main trends in qauntum software testing regarding the use of statistics based on repeated measurements and Hoare-like logics.\\ 
    \midrule 
    ~\cite{rw-Leite-testing_debugging}
    & 2025
    & Unspec. 
    & Unspec.
    & Software testing and debugging                   
    & Examined existing concepts and techniques for testing and debugging quantum software, and identified challenges and future directions.\\ 
    \midrule 
    ~\cite{rw-jim-qsw_dev}
    & 2024
    & Unspec.  
    & Unspec.
    & Software design and development 
    & Conducted a questionnaire-based survey to investigate modeling tools and quantum programming languages being used to design and develop quantum software.\\ 
    \midrule 
    ~\cite{rw-de-repository}
    & 2022
    & Unspec.  
    & Unspec.
    & Software design and development     
    & Summarized taxonomies of current usages and challenges perceived by quantum programmers.\\ 
    \midrule 
    ~\cite{rw-de-testing}     
    & 2022
    & Unspec. 
    & Unspec. 
    & Software testing and debugging
    & Examined the field of quantum software testing, with a particular focus on applying mutation-based testing techniques for quantum circuits. \\ 
    \midrule 
    ~\cite{rw-Zhao-qse}          
    & 2020 
    & Unspec.    
    & Unspec. 
    & Requirements, design, implementation, testing and maintenance                                          
    & Examined existing support for quantum software development\\ 
    \bottomrule 
\end{tabular} 

%% file: tables/related-opt.tex
\begin{tabular}{l p{0.1\textwidth} p{0.22\textwidth} p{0.35\textwidth} p{0.38\textwidth}} 
    \toprule 
    Works & Publication Year & Optimization Type  & Algorithms and Approaches  & Research Method and Scope \\ 
    \midrule 
    ~\cite{rw-abbas-quantum_opt}                   
    & 2024  
    & Quantum optimization                                           
    & Grover search, quantum adiabatic algorithm, quantum phase estimation, Gibbs sampling, approximation algorithms, variational methods
    & Via a systematic mapping study, overviewed existing quantum optimization algorithms, including their approximate and heuristic settings, core building blocks, and metrics for benchmarking. \\
    \midrule 
    ~\cite{rw-blekos-qaoa}                   
    & 2024  
    & Quantum optimization                                    
    & QAOA and its variants 
    & Analyzed the performance of QAOA in diverse application contexts, discussed hardware-specific challenges, and empirically compared QAOA variants. \\ 
    \midrule 
    ~\cite{rw-gemeinhardt-quantum_combinatorial_opt}                   
    & 2023 
    & Quantum optimization                                           
    & QAOA/VQE class, QA class, other class (e.g., Grover-based optimization algorithms, iterative quantum assisted eigensolvers), hybrid quantum-classical approaches 
    & Via SLR, investigated problem formulations, existing solutions, typical use cases, and research interest for solving combinatorial optimization problems with NISQ devices.\\ 
    \midrule 
    ~\cite{rw-albash-adiabatic}               
    & 2018  
    & Quantum optimization                                           
    & Quantum annealing, adiabatic Grover search algorithm, adiabatic Deutsch-Jozsa algorithm, adiabatic Bernstein-Vazirani algorithm, adiabatic PageRank algorithm, stochastic adiabatic quantum computation   
    & Specific to adiabatic quantum computing, discussed its underlying principles, algorithmic accomplishments and limitations, and 
    its role within the broader field of computational complexity theory.\\ 
    \midrule 
    ~\cite{rw-pooja-quantum_inpired}  
    & 2024  
    & Quantum-inspired optimization with general hardware                                           
    & Quantum-inspired genetic algorithms, quantum-inspired evolutionary algorithms, quantum-inspired PSO algorithms, quantum-inspired algorithms based on physics and chemistry laws 
    & Via an SLR, identified emerging research areas for quantum-inspired methods and their popularity, and discussed the hybridization of quantum-inspired algorithms with machine learning. \\ 
    \midrule 
    ~\cite{rw-ross-quantum_inspired} 
    & 2019                  
    & Quantum-inspired optimization with general hardware
    & Quantum-inspired evolutionary algorithms, quantum swarm evolutionary algorithms,  multi-objective quantum-nature-inspired algorithms,  quantum-inspired gravitation search algorithms                 
    & Analyzed the quantum chromosome encoding of the discussed quantum-inspired algorithms and discussed possible metaheuristics yielding the best results. \\ 
    \midrule 
    ~\cite{rw-zhang-qi_evol}                   
    & 2011  
    & Quantum-inspired optimization with general hardware
    & Binary observation quantum-inspired evolutionary algorithm and its three variants                                            
    & Discussed qubit representation, basic structure, and candidate optimization problems to be solved for the involved quantum-inspired algorithms, and presented a small-scale empirical study for  
    comparison between state-of-the-art evolutionary algorithms and quantum-inspired evolutionary algorithm variants. \\  
    \bottomrule 
\end{tabular} 

%% file: related.tex
In this section, we present the literature that is related to our SLR, including reviews, surveys, and SLRs conducted by the QSE community and beyond. 

\subsection{Within QSE}
QSE is defined as "an interdisciplinary field that focuses on the principles, methodologies, standards, and tools for designing, developing, testing, maintaining, and managing quantum software systems" and "the application of quantum-based techniques, including quantum algorithms such as quantum annealing, to solve complex problems in both classic and quantum software engineering"~\cite{TOSEMRoadmap2025}. 
By keeping this definition in mind, we examined the literature, identified reviews or surveys that deliver relatively comprehensive and meaningful results, and summary them in 
Table~\ref{tab:rw-qse}. From the table, we can observe that five out of the 10 related works are SLRs. These five papers include papers published until 2024. Our SLR, however, covers until early 2025 and is also concerned with different scopes. 

As for the scope of covering the phases of developing quantum software, the three
 reviews~\cite{rw-Zhao-qse, rw-Murillo-qse,rw-mandal-qse_qml} offer an overview of potential activities across the hypothesized quantum software development life cycle (SDLC). 
Some works focus on one or two specific phases of SDLC such as requirements engineering~\cite{rw-Sep-requirements}, software development~\cite{rw-jim-qsw_dev, rw-de-repository}, and testing and debugging~\cite{rw-Leite-testing_debugging, rw-Paltenghi-testing_analysis, rw-Garcia-testing, rw-de-testing}. In particular, testing and debugging of quantum software have garnered the most significant attention in the QSE community. In comparison, our survey does not target the activities within the quantum SDLC but explores quantum-based techniques for classical software engineering. 

Mandal et al.~\cite{rw-mandal-qse_qml} presented a comprehensive SLR, which is the most relevant to our study. Their SLR summarises and advocates the potential of quantum computing in software engineering, with the consideration of both quantum machine learning and quantum optimization. As concluded in their SLR, quantum machine learning has been used to detect software vulnerabilities and code smells while quantum optimization algorithms are promising in solving graph-based combinatorial optimization problems, a category into which certain software engineering optimization problems can be formulated. However, as emphasized in their SLR, the application of these algorithms in the software domain remains relatively unexplored. Compared to 
their work investigating learn-based methods, our SLR takes the perspective of SBSE, and zooms into algorithms for optimization, which either draw on the quantum mechanical principles for classical optimizers or directly exploit the power of real quantum hardware. To gain a comprehensive and systematic view of the research field, we followed the widely recognized software engineering body of knowledge~\cite{swebok2024} to guide the paper collection and analysis. This can ensure to achieve a more comprehensive state of the art and stay focused on software engineering optimization challenges.
 

\subsection{Beyond QSE}
Beyond QSE, there are related works from the quantum computing and artificial intelligence communities. In Table~\ref{tab:rw-opt}, we present several comprehensive surveys or review articles on quantum-based optimization algorithms, and summarize aspects that may be of interest to the QSE community. As shown in the table, for works on quantum optimizations~\cite{rw-abbas-quantum_opt, rw-blekos-qaoa, rw-gemeinhardt-quantum_combinatorial_opt, rw-albash-adiabatic}, QAOA and QA (or quantum adiabatic algorithm) have drawn great attention in academia. Meanwhile, it can be observed that
the works on quantum-inspired optimizations~\cite{rw-pooja-quantum_inpired, rw-ross-quantum_inspired, rw-zhang-qi_evol} are all about performing optimizations on general
 classical hardware (e.g., CPU), while quantum-inspired algorithms with specialized hardware are scarcely discussed in the literature. 
 

As for the applications, most evaluations presented in the related works are based on benchmark functions or problems, thereby a requirement to investigate the application of these algorithms to scalable and real-world problems is not satisfied. Practically, several review papers reported the feasibility of either quantum or quantum-inspired optimizations in areas like Industry 4.0~\cite{rw-sood-qi_industry4.0}, unmanned aerial vehicles~\cite{rw-kumar-quantum_drone}, and smart grids~\cite{rw-ullah-quantum_smartgrid}. Moreover, Mandal et al.~\cite{rw-mandal-feature} reviewed feature subset selection for pattern recognition applications based on both quantum and quantum-inspired optimization algorithms. Nevertheless, 
there are scarce reviews or surveys for applications of quantum or quantum-inspired optimization algorithms, particularly in addressing software engineering problems.


%% file: appendix.tex
\section*{Appendix}

\begin{table}[thp]
	\small
	\centering
	\caption{Formulated search queries for each database with corresponding number of found papers. Note that \textit{Total} refers to the number of unique found papers by removing duplicates across databases.}
	\label{tab:queries}
	\resizebox{.89\textwidth}{!}{
		\input{tables/queries.tex}
	}
\end{table}

\begin{table}
	\small
	\centering
	\caption{Summary of quantum solutions for addressing each kind of SE problem}
	\label{tab:rq_solution_summary_bySEproblem}
	\resizebox{.72\textwidth}{!}{
		\input{generated_files/RQ15_solution_summary_bySEproblem.tex}
	}
\end{table}

\begin{table}
	\small
	\centering
	\caption{Summary of quantum solutions for addressing each kind of problem}
	\label{tab:rq_solution_summary_byproblem}
	\resizebox{.99\textwidth}{!}{
		\input{generated_files/RQ15_solution_summary_byproblem.tex}
	}
\end{table}

%% file: tables/queries.tex
\begin{tabular} { l p{0.7\textwidth} r}\\ 
	\toprule 
	Name &  Search Query & \# of Found Papers\\ 
	
	\midrule 
	\dbIEEE & ("Abstract":quantum) AND
	("Abstract":minimi* OR "Abstract":optimi* OR "Abstract":prioriti* OR "Abstract":selection OR "Abstract":reduction) AND
	("Abstract":algorithm OR "Abstract":heuristic* OR "Abstract":search OR "Abstract":learning OR "Abstract":artificial intelligence OR "Abstract":ai) AND
	("Abstract":api OR "Abstract":develop* OR "Abstract":bug OR "Abstract":code OR "Abstract":coding OR "Abstract":debug OR "Abstract":defect OR "Abstract":deploy OR "Abstract":evolution OR "Abstract":fault OR "Abstract":fix OR "Abstract":maintenance OR "Abstract":program OR "Abstract":refactor* OR "Abstract":repair OR "Abstract":requirement OR "Abstract":test* OR "Abstract":verification OR "Abstract":validation OR "Abstract":vulnerab* OR "Abstract":configur*) AND
	("Full Text Only":software)  &  557\\ 
	
	\midrule

	\dbACM & [Abstract: quantum] AND 
	[[Abstract: minimi*] OR [Abstract: optimi*] OR [Abstract: prioriti*] OR [Abstract: selection] OR [Abstract: reduction]] AND 
	[[Abstract: algorithm] OR [Abstract: heuristic*] OR [Abstract: search] OR [Abstract: learning] OR [Abstract: artificial intelligence] OR [Abstract: ai]] AND 
	[[Abstract: api] OR [Abstract: develop*] OR [Abstract: bug] OR [Abstract: code] OR [Abstract: coding] OR [Abstract: debug] OR [Abstract: defect] OR [Abstract: deploy] OR [Abstract: evolution] OR [Abstract: fault] OR [Abstract: fix] OR [Abstract: maintenance] OR [Abstract: program] OR [Abstract: refactor*] OR [Abstract: repair] OR [Abstract: requirement] OR [Abstract: test*] OR [Abstract: verification] OR [Abstract: validation] OR [Abstract: vulnerab*] OR [Abstract: configur*]] AND 
	[Full Text: software] & 222 \\ 
	
	\midrule 
	
	\dbScopus & ( ABS ( ( quantum ) AND 
	( minimi* OR optimi* OR prioriti* OR selection OR reduction ) AND ( algorithm OR heuristic* OR search OR learning OR artificial AND intelligence OR ai ) AND 
	( api OR develop* OR bug OR code OR coding OR debug OR defect OR deploy OR evolution OR fault OR fix OR maintenance OR program OR refactor* OR repair OR requirement OR test* OR verification OR validation OR vulnerab* OR configur* ) ) AND 
	ALL ( software ) ) AND 
	PUBYEAR > 2013 AND PUBYEAR < 2026 AND ( LIMIT-TO ( SUBJAREA , "COMP" ) ) AND ( LIMIT-TO ( DOCTYPE , "ar" ) OR LIMIT-TO ( DOCTYPE , "cp" ) OR LIMIT-TO ( DOCTYPE , "re" ) ) AND ( LIMIT-TO ( LANGUAGE , "English" ) ) AND ( LIMIT-TO ( SRCTYPE , "j" ) OR LIMIT-TO ( SRCTYPE , "p" ) ) & 68 \\
	
	\midrule 
	
	\dbWiley & "( quantum ) AND 
	( minimi* OR optimi* OR prioriti* OR selection OR reduction ) AND 
	( algorithm OR heuristic* OR search OR learning OR artificial AND intelligence OR ai ) AND 
	( api OR develop* OR bug OR code OR coding OR debug OR defect OR deploy OR evolution OR fault OR fix OR maintenance OR program OR refactor* OR repair OR requirement OR test* OR verification OR validation OR vulnerab* OR configur* )" in Abstract And 
	"software" anywhere & 292 \\
	
	\midrule 
	
	\dbWoS & AB=(( quantum ) AND 
	( minimi* OR optimi* OR prioriti* OR selection OR reduction ) AND 
	( algorithm OR heuristic* OR search OR learning OR artificial AND intelligence OR ai ) AND 
	( api OR develop* OR bug OR code OR coding OR debug OR defect OR deploy OR evolution OR fault OR fix OR maintenance OR program OR refactor* OR repair OR requirement OR test* OR verification OR validation OR vulnerab* OR configur* ) ) AND 
	ALL=(software) & 268 \\
	
	\midrule 
	\dbSpringer & ( quantum ) AND 
	( minimi* OR optimi* OR prioriti* OR selection OR reduction ) AND 
	( algorithm OR heuristic* OR search OR learning OR artificial intelligence OR ai ) AND 
	( api OR develop* OR bug OR code OR coding OR debug OR defect OR deploy OR evolution OR fault OR fix OR maintenance OR program OR refactor* OR repair OR requirement OR test* OR verification OR validation OR vulnerab* OR configur* ) AND 
	software & 779 \\
	\midrule 
	\textit{Sum} &  & 2186\\
	\textit{Total\textsuperscript{*}} &  & 2083\\
	\bottomrule 
\end{tabular}  

%% file: generated_files/RQ15_solution_summary_bySEproblem.tex
\begin{tabular} { l l l l} \\ 
\toprule 
 SE Problem & Objectives and Constraints & Solutions & Papers\\ 
\midrule  
\emph{scheduling problem}&Many $\rightarrow$ Many + Y & Quantum-inspired: PSO + Circuit-based & \cite{ \paperseventysecond }\\ 
&Multiple $\rightarrow$ Single + Y & Quantum-inspired: GA + Physics-based & \cite{ \paperthird }\\ 
& & Quantum-inspired: PSO + Circuit-based & \cite{ \papereightysecond }\\ 
& & Quantum-inspired: PSO + Physics-based & \cite{ \paperthird }\\ 
& & Quantum-inspired: QA + Physics-based & \cite{ \paperfortyeighth }\\ 
& & Quantum-inspired: SSA + Physics-based & \cite{ \paperfortyfirst }\\ 
&Single $\rightarrow$ Single + Y & Quantum-inspired: Un-specified + Circuit-based & \cite{ \paperninetyfirst }\\ 
\midrule  
\emph{test suite minimization}&Multiple $\rightarrow$ Single + N & Hybrid: QA + D-Wave + QUBO + QH & \cite{ \papersixtyfourth }\\ 
& & Hybrid: QAOA + Qiskit + Ising model + ICS/NCS/QH & \cite{ \paperthirteenth }\\ 
&Single $\rightarrow$ Single + N & Quantum-inspired: ACO + Circuit-based & \cite{ \papersixteenth }\\ 
& & Quantum-inspired: BA + Physics-based & \cite{ \paperthirtythird }\\ 
& & Quantum-inspired: GA + Physics-based & \cite{ \papertenth }\\ 
& & Quantum-inspired: PSO + Physics-based & \cite{ \papertenth,\paperthirtyeighth }\\ 
& & Quantum-inspired: Un-specified + Physics-based & \cite{ \paperseventyfourth }\\ 
&Single $\rightarrow$ Single + Y & Quantum-inspired: GA + Circuit-based & \cite{ \paperseventyseventh }\\ 
& & Quantum: Grover + Un-specified + None + Un-specified & \cite{ \paperthirtyseventh }\\ 
\midrule  
\emph{software failure prediction}&Single $\rightarrow$ Single + N & Quantum-inspired: CSO + Un-specified & \cite{ \papertwelfth }\\ 
& & Quantum-inspired: GSA + Circuit-based & \cite{ \papersixtyseventh,\paperseventythird }\\ 
& & Quantum-inspired: PSO + Physics-based & \cite{ \paperthirtieth,\paperseventyfirst }\\ 
& & Quantum: QA + D-Wave + QUBO + QH & \cite{ \papereleventh }\\ 
&Single $\rightarrow$ Single + Y & Quantum-inspired: PSO + Un-specified & \cite{ \papertwentysixth }\\ 
\midrule  
\emph{join order problem}&Single $\rightarrow$ Single + Y & Quantum-inspired: QA + Physics-based & \cite{ \paperfifth }\\ 
& & Hybrid: QAOA + Qiskit + QUBO + CS & \cite{ \paperfiftyfourth }\\ 
& & Hybrid: QAOA + Qiskit + QUBO + QH & \cite{ \paperninetythird }\\ 
& & Hybrid: VQE + Qiskit + QUBO + CS & \cite{ \paperfiftyfourth }\\ 
& & Quantum: QA + D-Wave + CQM + QH & \cite{ \paperfiftieth }\\ 
& & Quantum: QA + D-Wave + HUBO + QH & \cite{ \paperfiftysecond }\\ 
& & Quantum: QA + D-Wave + QUBO + QH & \cite{ \paperfiftieth,\paperfiftysecond,\paperfiftyfourth,\paperninetythird }\\ 
\midrule  
\emph{security attack identification}&Single $\rightarrow$ Single + N & Quantum-inspired: ACO + Circuit-based & \cite{ \paperfortysecond }\\ 
& & Quantum-inspired: ALO + Circuit-based & \cite{ \papertwentysecond }\\ 
& & Hybrid: QA + D-Wave + QUBO + QH & \cite{ \paperfiftyfifth }\\ 
&Single $\rightarrow$ Single + Y & Hybrid: QA + D-Wave + Un-specified + QH & \cite{ \paperfiftysixth }\\ 
\midrule  
\emph{covering array generation}&Single $\rightarrow$ Single + N & Quantum-inspired: EA + Circuit-based & \cite{ \papersixtysixth }\\ 
& & Quantum-inspired: PSO + Circuit-based & \cite{ \papertwentieth }\\ 
& & Quantum-inspired: PSO + Physics-based & \cite{ \paperseventysixth }\\ 
&Single $\rightarrow$ Single + Y & Quantum-inspired: PSO + Physics-based & \cite{ \paperthirtyfourth,\paperseventyfifth }\\ 
\midrule  
\emph{allocation problem}&Multiple $\rightarrow$ Single + N & Quantum-inspired: PSO + Physics-based & \cite{ \paperseventieth }\\ 
&Multiple $\rightarrow$ Single + Y & Quantum-inspired: Un-specified + Circuit-based & \cite{ \papereightysixth }\\ 
& & Hybrid: QA + D-Wave + QUBO + QH & \cite{ \papereightyseventh,\papereightyeighth }\\ 
\midrule  
\emph{controller placement}&Multiple $\rightarrow$ Multiple + N & Quantum-inspired: SSA + Physics-based & \cite{ \papereightythird }\\ 
&Multiple $\rightarrow$ Multiple + Y & Quantum-inspired: SSA + Physics-based & \cite{ \paperfortyfourth }\\ 
&Single $\rightarrow$ Single + Y & Quantum-inspired: PSO + Physics-based & \cite{ \paperthirtyfirst }\\ 
\midrule  
\emph{routing optimization}&Many $\rightarrow$ Single + Y & Quantum-inspired: ROA + Circuit-based & \cite{ \papereightyninth }\\ 
&Multiple $\rightarrow$ Multiple + Y & Hybrid: QAOA + Qiskit + Ising model + QH & \cite{ \papereightyfourth }\\ 
& & Hybrid: QAOA + Qiskit + QUBO + QH & \cite{ \papereightyfourth }\\ 
&Multiple $\rightarrow$ Single + N & Quantum-inspired: EA + Circuit-based & \cite{ \paperninetysecond }\\ 
&Single $\rightarrow$ Single + Y & Hybrid: QAOA + Qiskit + Ising model + QH & \cite{ \papereightyfourth }\\ 
& & Hybrid: QAOA + Qiskit + QUBO + QH & \cite{ \papereightyfourth }\\ 
\midrule  
\emph{test case prioritization}&Single $\rightarrow$ Single + N & Quantum-inspired: BA + Physics-based & \cite{ \paperthirtythird }\\ 
& & Quantum-inspired: GA + Physics-based & \cite{ \papertenth }\\ 
& & Quantum-inspired: PSO + Physics-based & \cite{ \papertenth,\paperthirtyeighth }\\ 
\midrule  
\emph{test case selection}&Multiple $\rightarrow$ Single + N & Hybrid: QAOA + Qiskit + Ising model + ICS/NCS/QH & \cite{ \paperthirteenth }\\ 
&Multiple $\rightarrow$ Single + Y & Hybrid: QA + D-Wave + QUBO + QH & \cite{ \papersixtythird }\\ 
&Single $\rightarrow$ Single + N & Quantum-inspired: PSO + Physics-based & \cite{ \paperthirtyeighth }\\ 
\midrule  
\emph{multiple query optimization}&Single $\rightarrow$ Single + Y & Hybrid: QAOA + Qiskit + QUBO + CS/QH & \cite{ \paperfiftyeighth }\\ 
& & Quantum: QA + D-Wave + QUBO + QH & \cite{ \paperfiftythird }\\ 
\midrule  
\emph{security management}&Multiple $\rightarrow$ Single + Y & Quantum: QA + D-Wave + QUBO + QH & \cite{ \papersixtyeighth }\\ 
&Single $\rightarrow$ Single + Y & Quantum: QA + D-Wave + QUBO + QH & \cite{ \papertwentyfifth }\\ 
\midrule  
\emph{service deployment}&Single $\rightarrow$ Single + Y & Quantum-inspired: GA + Circuit-based & \cite{ \paperthirtysixth,\paperthirtyninth }\\ 
\midrule  
\emph{software requirement selection}&Multiple $\rightarrow$ Multiple + Y & Quantum-inspired: AVOA + Circuit-based & \cite{ \papertwentyninth }\\ 
& & Quantum-inspired: EA + Circuit-based & \cite{ \papertwentyeighth }\\ 
\midrule  
\emph{clustering protocol optimization}&Many $\rightarrow$ Single + Y & Quantum-inspired: WOA + Circuit-based & \cite{ \papereightyfifth }\\ 
\midrule  
\emph{code clone detection}&Single $\rightarrow$ Single + Y & Quantum: QA + D-Wave + QUBO + QH & \cite{ \paperfirst }\\ 
& & Quantum: QA + D-Wave + QUDO + QH & \cite{ \paperfirst }\\ 
\midrule  
\emph{configuration prioritization}&Single $\rightarrow$ Single + Y & Hybrid: QAOA + Qiskit + HUBO + ICS & \cite{ \papersecond }\\ 
& & Hybrid: QAOA + Qiskit + Ising model + ICS & \cite{ \papersecond }\\ 
& & Hybrid: QAOA + Qiskit + QUBO + ICS & \cite{ \papersecond }\\ 
\midrule  
\emph{configuration selection}&Single $\rightarrow$ Single + Y & Hybrid: QAOA + Qiskit + HUBO + ICS & \cite{ \papersecond }\\ 
& & Hybrid: QAOA + Qiskit + Ising model + ICS & \cite{ \papersecond }\\ 
& & Hybrid: QAOA + Qiskit + QUBO + ICS & \cite{ \papersecond }\\ 
\midrule  
\emph{database index selection}&Single $\rightarrow$ Single + Y & Hybrid: QA + D-Wave + QUBO + QH & \cite{ \papersixtieth }\\ 
\midrule  
\emph{feature selection}&Single $\rightarrow$ Single + Y & Quantum: QA + D-Wave + QUBO + QH & \cite{ \paperninetieth }\\ 
\midrule  
\emph{network delay optimization}&Single $\rightarrow$ Single + Y & Quantum-inspired: GA + Circuit-based & \cite{ \papereightieth }\\ 
\midrule  
\emph{node localization}&Single $\rightarrow$ Single + N & Quantum-inspired: SSA + Un-specified & \cite{ \papertwentythird }\\ 
\midrule  
\emph{queuing delay optimization}&Multiple $\rightarrow$ Single + Y & Quantum-inspired: GA + Circuit-based & \cite{ \paperseventyninth }\\ 
\midrule  
\emph{resource optimization}&Many $\rightarrow$ Single + Y & Quantum-inspired: Un-specified + Circuit-based & \cite{ \paperfortyfifth }\\ 
\midrule  
\emph{service configuration}&Multiple $\rightarrow$ Single + Y & Quantum-inspired: GA + Circuit-based & \cite{ \paperfortieth }\\ 
\midrule  
\emph{software reliability growth model}&Single $\rightarrow$ Single + N & Quantum-inspired: PSO + Physics-based & \cite{ \paperthirtysecond }\\ 
\midrule  
\emph{wcet evaluation}&Single $\rightarrow$ Single + Y & Hybrid: QAOA + Qiskit + QUBO + CS & \cite{ \papertwentyseventh }\\ 
& & Quantum: QA + D-Wave + QUBO + QH & \cite{ \papertwentyseventh }\\ 
\bottomrule 
\end{tabular} 

%% file: generated_files/RQ15_solution_summary_byproblem.tex
\begin{tabular} { l l p{0.5\textwidth} p{0.12\textwidth}} \\ 
\toprule 
 Objectives and Constraints & Solutions  & SE Problem & Papers\\ 
\midrule  
Many $\rightarrow$ Many + Y & Quantum-inspired: PSO + Circuit-based & scheduling problem & \cite{ \paperseventysecond }\\ 
\midrule  
Many $\rightarrow$ Single + Y & Quantum-inspired: ROA + Circuit-based & routing optimization & \cite{ \papereightyninth }\\ 
 & Quantum-inspired: Un-specified + Circuit-based & resource optimization & \cite{ \paperfortyfifth }\\ 
 & Quantum-inspired: WOA + Circuit-based & clustering protocol optimization & \cite{ \papereightyfifth }\\ 
\midrule  
Multiple $\rightarrow$ Multiple + N & Quantum-inspired: SSA + Physics-based & controller placement & \cite{ \papereightythird }\\ 
\midrule  
Multiple $\rightarrow$ Multiple + Y & Quantum-inspired: AVOA + Circuit-based & software requirement selection & \cite{ \papertwentyninth }\\ 
 & Quantum-inspired: EA + Circuit-based & software requirement selection & \cite{ \papertwentyeighth }\\ 
 & Quantum-inspired: SSA + Physics-based & controller placement & \cite{ \paperfortyfourth }\\ 
 & Hybrid: QAOA + Qiskit + Ising model + QH & routing optimization & \cite{ \papereightyfourth }\\ 
 & Hybrid: QAOA + Qiskit + QUBO + QH & routing optimization & \cite{ \papereightyfourth }\\ 
\midrule  
Multiple $\rightarrow$ Single + N & Quantum-inspired: EA + Circuit-based & routing optimization & \cite{ \paperninetysecond }\\ 
 & Quantum-inspired: PSO + Physics-based & allocation problem & \cite{ \paperseventieth }\\ 
 & Hybrid: QA + D-Wave + QUBO + QH & test suite minimization & \cite{ \papersixtyfourth }\\ 
 & Hybrid: QAOA + Qiskit + Ising model + ICS/NCS/QH & test case selection, test suite minimization & \cite{ \paperthirteenth }\\ 
\midrule  
Multiple $\rightarrow$ Single + Y & Quantum-inspired: GA + Circuit-based & service configuration, queuing delay optimization & \cite{ \paperfortieth,\paperseventyninth }\\ 
 & Quantum-inspired: GA + Physics-based & scheduling problem & \cite{ \paperthird }\\ 
 & Quantum-inspired: PSO + Circuit-based & scheduling problem & \cite{ \papereightysecond }\\ 
 & Quantum-inspired: PSO + Physics-based & scheduling problem & \cite{ \paperthird }\\ 
 & Quantum-inspired: QA + Physics-based & scheduling problem & \cite{ \paperfortyeighth }\\ 
 & Quantum-inspired: SSA + Physics-based & scheduling problem & \cite{ \paperfortyfirst }\\ 
 & Quantum-inspired: Un-specified + Circuit-based & allocation problem & \cite{ \papereightysixth }\\ 
 & Hybrid: QA + D-Wave + QUBO + QH & test case selection, allocation problem, allocation problem & \cite{ \papersixtythird,\papereightyseventh,\papereightyeighth }\\ 
 & Quantum: QA + D-Wave + QUBO + QH & security management & \cite{ \papersixtyeighth }\\ 
\midrule  
Single $\rightarrow$ Single + N & Quantum-inspired: ACO + Circuit-based & test suite minimization, security attack identification & \cite{ \papersixteenth,\paperfortysecond }\\ 
 & Quantum-inspired: ALO + Circuit-based & security attack identification & \cite{ \papertwentysecond }\\ 
 & Quantum-inspired: BA + Physics-based & test case prioritization, test suite minimization & \cite{ \paperthirtythird }\\ 
 & Quantum-inspired: CSO + Un-specified & software failure prediction & \cite{ \papertwelfth }\\ 
 & Quantum-inspired: EA + Circuit-based & covering array generation & \cite{ \papersixtysixth }\\ 
 & Quantum-inspired: GA + Physics-based & test suite minimization, test case prioritization & \cite{ \papertenth }\\ 
 & Quantum-inspired: GSA + Circuit-based & software failure prediction, software failure prediction & \cite{ \papersixtyseventh,\paperseventythird }\\ 
 & Quantum-inspired: PSO + Circuit-based & covering array generation & \cite{ \papertwentieth }\\ 
 & Quantum-inspired: PSO + Physics-based & test suite minimization, test case prioritization, software failure prediction, software reliability growth model, test case prioritization, test case selection, test suite minimization, software failure prediction, covering array generation & \cite{ \papertenth,\paperthirtieth,\paperthirtysecond,\paperthirtyeighth,\paperseventyfirst,\paperseventysixth }\\ 
 & Quantum-inspired: SSA + Un-specified & node localization & \cite{ \papertwentythird }\\ 
 & Quantum-inspired: Un-specified + Physics-based & test suite minimization & \cite{ \paperseventyfourth }\\ 
 & Hybrid: QA + D-Wave + QUBO + QH & security attack identification & \cite{ \paperfiftyfifth }\\ 
 & Quantum: QA + D-Wave + QUBO + QH & software failure prediction & \cite{ \papereleventh }\\ 
\midrule  
Single $\rightarrow$ Single + Y & Quantum-inspired: GA + Circuit-based & service deployment, service deployment, test suite minimization, network delay optimization & \cite{ \paperthirtysixth,\paperthirtyninth,\paperseventyseventh,\papereightieth }\\ 
 & Quantum-inspired: PSO + Physics-based & controller placement, covering array generation, covering array generation & \cite{ \paperthirtyfirst,\paperthirtyfourth,\paperseventyfifth }\\ 
 & Quantum-inspired: PSO + Un-specified & software failure prediction & \cite{ \papertwentysixth }\\ 
 & Quantum-inspired: QA + Physics-based & join order problem & \cite{ \paperfifth }\\ 
 & Quantum-inspired: Un-specified + Circuit-based & scheduling problem & \cite{ \paperninetyfirst }\\ 
 & Hybrid: QA + D-Wave + QUBO + QH & database index selection & \cite{ \papersixtieth }\\ 
 & Hybrid: QA + D-Wave + Un-specified + QH & security attack identification & \cite{ \paperfiftysixth }\\ 
 & Hybrid: QAOA + Qiskit + HUBO + ICS & configuration selection, configuration prioritization & \cite{ \papersecond }\\ 
 & Hybrid: QAOA + Qiskit + Ising model + ICS & configuration selection, configuration prioritization & \cite{ \papersecond }\\ 
 & Hybrid: QAOA + Qiskit + Ising model + QH & routing optimization & \cite{ \papereightyfourth }\\ 
 & Hybrid: QAOA + Qiskit + QUBO + CS & wcet evaluation, join order problem & \cite{ \papertwentyseventh,\paperfiftyfourth }\\ 
 & Hybrid: QAOA + Qiskit + QUBO + CS/QH & multiple query optimization & \cite{ \paperfiftyeighth }\\ 
 & Hybrid: QAOA + Qiskit + QUBO + ICS & configuration selection, configuration prioritization & \cite{ \papersecond }\\ 
 & Hybrid: QAOA + Qiskit + QUBO + QH & routing optimization, join order problem & \cite{ \papereightyfourth,\paperninetythird }\\ 
 & Hybrid: VQE + Qiskit + QUBO + CS & join order problem & \cite{ \paperfiftyfourth }\\ 
 & Quantum: Grover + Un-specified + None + Un-specified & test suite minimization & \cite{ \paperthirtyseventh }\\ 
 & Quantum: QA + D-Wave + CQM + QH & join order problem & \cite{ \paperfiftieth }\\ 
 & Quantum: QA + D-Wave + HUBO + QH & join order problem & \cite{ \paperfiftysecond }\\ 
 & Quantum: QA + D-Wave + QUBO + QH & code clone detection, security management, wcet evaluation, join order problem, join order problem, multiple query optimization, join order problem, feature selection, join order problem & \cite{ \paperfirst,\papertwentyfifth,\papertwentyseventh,\paperfiftieth,\paperfiftysecond,\paperfiftythird,\paperfiftyfourth,\paperninetieth,\paperninetythird }\\ 
 & Quantum: QA + D-Wave + QUDO + QH & code clone detection & \cite{ \paperfirst }\\ 
\bottomrule 
\end{tabular} 